\documentclass[a4paper,fleqn]{cas-sc}
\usepackage{float}
\usepackage[numbers]{natbib}

\graphicspath{{figures/}}

\usepackage{multirow}
\usepackage{physics,bbm,mathtools,amsmath,amsthm,amssymb,amsfonts}
\usepackage{booktabs}
\usepackage{xcolor}
\usepackage{amsmath}
\usepackage{array}

\usepackage{algorithmic}
\usepackage{algorithm}

\usepackage{pmboxdraw}

\DeclareMathOperator*{\argmax}{arg\,max}

\begin{document}
\let\WriteBookmarks\relax
\def\floatpagepagefraction{1}
\def\textpagefraction{.001}

\shorttitle{}

\shortauthors{ }

\title [mode = title]{Benchmarking multi-component signal processing methods in the time-frequency plane}                      



\author[1,2]{J. M. Miramont}[orcid=0000-0002-3847-7811]

\cormark[1]


\ead{jmiramontt@univ-lille.fr}


\credit{Conceptualization of this study, Methodology, Software}

\affiliation[1]{organization={Universit\'e de Lille, CNRS, Centrale Lille, UMR 9189 Centre de Recherche en Informatique, Signal et Automatique de Lille (CRIStAL)},
    city={Lille},
    postcode={F-59000}, 
    country={France}}

\author[1]{R. Bardenet}[orcid=0000-0002-1094-9493]

\author[1]{P. Chainais}[orcid=0000-0003-4377-7584]


\credit{Data curation, Writing - Original draft preparation}

\affiliation[2]{organization={Nantes Université, UR 4642 Institut de Recherche en Énergie Électrique de Nantes Atlantique (IREENA)},
    city={Saint-Nazaire},
    postcode={F-44600}, 
    country={France}}

\author[2]{F. Auger}[orcid=0000-0001-9158-1784]





\begin{abstract}
    Signal processing in the time-frequency plane has a long history and remains a field of methodological innovation.
    For instance, detection and denoising based on the zeros of the spectrogram have been proposed since 2015, contrasting with a long history of focusing on larger values of the spectrogram.
    Yet, unlike neighboring fields like optimization and machine learning, time-frequency signal processing lacks widely-adopted benchmarking tools.
    In this work, we contribute an open-source, Python-based toolbox termed MCSM-Benchs for benchmarking multi-component signal analysis methods, and we demonstrate our toolbox on three time-frequency benchmarks.
    First, we compare different methods for signal detection based on the zeros of the spectrogram, including unexplored variations of previously proposed detection tests. 
    Second, we compare zero-based denoising methods to both classical and novel methods based on large values and ridges of the spectrogram. 
    Finally, we compare the denoising performance of these methods against typical spectrogram thresholding strategies, in terms of post-processing artifacts commonly referred to as musical noise.  
    At a low level, the obtained results provide new insight on the assessed approaches, and in particular research directions to further develop zero-based methods. 
    At a higher level, our benchmarks exemplify the benefits of using a public, collaborative, common framework for benchmarking.
\end{abstract}





\maketitle

\section{Introduction}

Time-frequency (TF) analysis is a vast area of research within signal processing. 
TF representations provide the means to deal with non-stationary signals, the frequency content of which varies with the time, helping to discern patterns that reveal the signal TF structure \cite{flandrin1998time}.   
Signal processing methods relying on TF representations, particularly on the well-known spectrogram, include the study of several high-level, complex problems like source separation \cite{villasana2023eusipco,sawada2019}, speech enhancement \cite{michelsanti2021overview} and music information processing and retrieval \cite{simonetta2019multimodal}. 
Methodological research also focuses on more elementary tasks, such as detecting an unknown signal in noise, \cite{bardenet2018zeros, bardenet2021time, ghosh2022signal, pascal2022covariant, pascalfamille, miramont2023unsupervised} and estimation of the signal (or its components) under stationary noise, \cite{meignen2016adaptive, harmouche2017sliding, meignen2018retrieval, laurent2020novel, colominas2020fully, laurent2021novel, legros2021novel, legros2022pb, ghosh2022signal, legros2022instantaneous, legros2022time} and new algorithms that tackle these elementary tasks keep appearing. 
In particular, recent times have seen the emergence of detection and denoising methods based on the \emph{zeros} of the spectrogram \cite{flandrin2015time, bardenet2018zeros, ghosh2022signal, miramont2023unsupervised}.

The use of the zeros of the spectrogram, rather than its more commonly-used \emph{large values}, and modulus maxima in particular, can be seen as a paradigm shift. 
Approaches based on zeros prompted further study of some theoretical aspects, \cite{bardenet2018zeros, ghosh2022signal, courbot2023sparse} and even fostered the interest in the zeros of different transforms like the continuous wavelet \cite{koliander2019filtering,bardenet2021time}, Kravchuk \cite{bardenet2021time,pascal2022covariant,pascalfamille} and Stockwell \cite{moukadem2021zeros} transforms.
Nevertheless, more practical aspects of the use of spectrogram zeros and their applications in signal detection and denoising are yet to be studied. 
For instance, the behavior of the zeros of the spectrogram in the presence of interference between signal components still comes with many open questions \cite{laurent2023novel}.
One of the motivations of this paper is to empirically investigate the pros and cons of current zero-based approaches.
The conditions under which approaches based on either zeros or large values of the spectrogram yield the best performance remain to be investigated. 

A systematic way of evaluating the advantages and limitations of computational methods is through \emph{benchmarks}, i.e. normalized comparisons of a wide array of existing methods on a predefined set of representative problems.
This is a standard procedure in many applied fields, like in numerical optimization \cite{hansen2021coco,bartz2020benchmarking} or machine learning \cite{benchopt, mattson2020mlperf}.
Such a benchmarking procedure is desirable in signal processing as well, where new methods also need to be thoroughly compared with existing approaches to delineate their relevance.
However, considering different scenarios for making these comparisons (i.e. the level and nature of the noise involved, the characteristics of the signals, etc.) can be difficult and time-consuming. 
Faced with this, researchers usually restrict themselves to a limited number of experiments in an effort to highlight the benefits of their approach.
This brings us to the second goal of this article, which is introducing \emph{MCSM-Benchs}, a Python-based toolbox for benchmarking multi-component signal processing methods.
MCSM-Benchs offers signal processing practitioners a common framework to easily carry out extensive contrasting between approaches in an objective way.

We shall mainly focus on signal detection and denoising methods throughout this article, but MCSM-Benchs also provides the means for testing methods for instantaneous frequency estimation, component retrieval and component number estimation \cite{miramont2023eusipco}. 

Three benchmarks are developed here in order to illustrate, on the one hand, the kind of evaluations that can be done with MCSM-Benchs, and on the other hand, how new insights can be gained from these comparisons. 
The benchmarks are freely available\footnote{\url{https://github.com/jmiramont/benchmarks-detection-denoising}} and they are planned to be regularly updated, adding new approaches and results.
Prospective users can either contribute their own methods to add them to the benchmarks, or download the toolbox to perform more personalized tests using templates, following the instructions and examples given in the repository.

In particular, the potential of signal processing methods based on the analysis of zeros of the spectrogram is studied thanks to the proposed benchmarking framework.
The first benchmark evaluates the performance of signal detection methods based on the zeros of the spectrogram, extending the results obtained in a previous work \cite{bardenet2018zeros}.

The second benchmark focuses on comparing methods coming from two paradigms: methods based on large values of the spectrogram and recent approaches that exploit the structure of the zeros of the spectrogram to obtain information about the signal.
This is the occasion to introduce a new strategy to automatically tune the parameters of zero-based methods for a given signal.
Finally, a third benchmark, including a real-world audio signal, compares post-processing quality after applying methods based on spectrogram zeros and spectrogram thresholding, considering the remaining \emph{musical noise} \cite{goh1998postprocessing}.

The article is organized as follows. Sec. \ref{sec:toolbox} outlines the main components of MCSM-Benchs.
Sec. \ref{sec:methods} defines the three benchmarks, as well as the signal denoising and detection methods used in each simulation. 
Sec. \ref{sec:results} is devoted to the analysis of results obtained in each benchmark. 
Sec. \ref{sec:discussion} offers an overall discussion of the results, along with some future works. 
Conclusions are then finally drawn in Sec. \ref{sec:conclusion}.

\section{MCSM-Benchs: a common framework for benchmarking multi-component signal processing methods} \label{sec:toolbox}
This section describes the main elements that constitute MCSM-Benchs, a toolbox for benchmarking multi-component signal based methods, and its capabilities.   

\subsection{Main components of MCSM-Benchs}
\begin{figure*}
 \includegraphics[width = \textwidth]{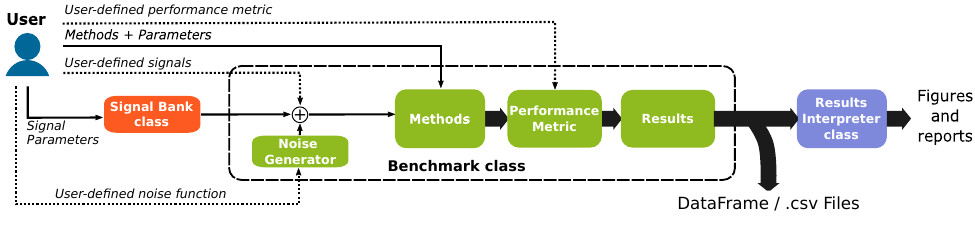}
 \caption{Block diagram describing the relationship between the elements of a typical benchmark created using MCSM-Benchs. Inward arrows represent the User inputs. Dotted lines represent optional inputs from the User.} \label{fig::block_diagram}
\end{figure*}

MCSM-Benchs was conceived with \emph{modularity} in mind, so that different modules can interact with each other; they may also be easily updated, or replaced, without affecting the other components in a benchmark. 
The object-oriented programming paradigm was extensively used, and the basic components of a benchmark are represented by different Python \emph{classes}. 
MCSM-Benchs mainly consists of three classes: 1) the \texttt{SignalBank} class, 2) the \texttt{Benchmark} class and 3) the \texttt{ResultsInterpreter} class.
Figure \ref{fig::block_diagram} shows a block diagram of a benchmark produced with the toolbox, where the interaction between the three main classes and a typical pipeline are shown. 
The inputs given by the user are a series of simulation parameters, one or several methods to be evaluated and, eventually, different sets of parameters for each method and performance metrics (optional input parameters are denoted with dotted lines in the figure).
The main simulation parameters include:
\begin{enumerate}
    \item The task (in this paper, either denoising or detection).
    \item The number of time samples of the signals.
    \item The Signal-to-Noise ratio (in dB): a tuple of values that determines the levels of noise used to contaminate the signals.
    \item The number of repetitions, i.e. how many noise realizations are used for each signal.
\end{enumerate}

Having set the benchmark parameters, independent noise realizations are added to the signals generated by the \texttt{SignalBank} class (or provided by the user) according to the given parameters. 
These are then fed to the methods by the \texttt{Benchmark} class, which encapsulates most of the pipeline.
It applies the methods to the noisy signals, compute the performance metric and, finally, gather all the results. 
The latter can then be analyzed by the \texttt{ResultsInterpreter} class, which generates output files, tables, and interactive figures.
Following sections briefly describe the three main Python classes and their components.

\subsubsection{The \texttt{SignalBank} class}
Noisy synthetic signals provide the means to quantitatively evaluate the performance of a technique, since both the noiseless version of the signal and the noise are known.
Moreover, the signals can be designed to pose specific challenges to the methods, which is relevant when the approaches to be benchmarked are based on \emph{models} of either the signal, the noise or both.

\begin{figure}
\centering
 \includegraphics[width = 0.5\textwidth]{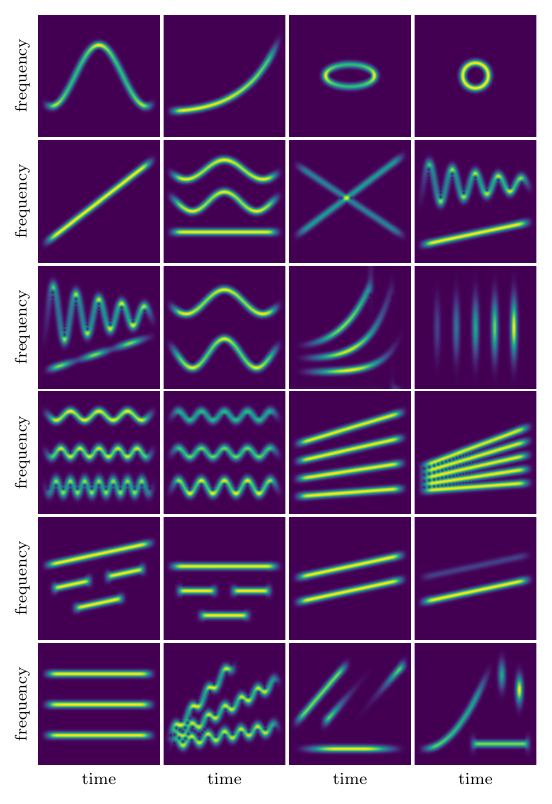}
 \caption{Spectrograms of some the signals available in the Signal Bank.} \label{fig::spectrograms}
\end{figure}

With this in mind, more than 20 synthetic signals with different time-frequency structures are offered to the user.
Figure \ref{fig::spectrograms} displays the spectrograms of some of the available signals.
\paragraph{The \texttt{Signal} class:}
Additionally, the \texttt{SignalBank} can output the signals as instances of a custom \texttt{Signal} class. 
In practice, these behave like regular arrays, but they also store useful information about the generated signal like:
\begin{itemize}
  \setlength\itemsep{0.1em}
    \item The number of signal components.
    \item The number of component at each time instant.
    \item The instantaneous frequency of each component.
\end{itemize} 
All this information can be queried by the methods and the performance functions, since they receive the signal as an input parameter.
This can be useful to create benchmarks on methods that estimate those quantities, or need estimates of them to work \cite{miramont2022public, miramont2023eusipco}.
For instance, the approaches described in \cite{legros2021novel, harmouche2017sliding} require the number of components, or the number of components for each time instant. 

Although the \texttt{SignalBank} can help standardize the benchmarks created with MCSM-Benchs, users are not limited to use the signals synthesized by this class, and can also use real-world signals when creating new benchmarks.

\subsubsection{The \texttt{Benchmark} class}
This class is the core of MCSM-Benchs, representing a benchmark itself.
It receives all the simulation parameters as inputs  and apply the methods to the signals.
Fig. \ref{fig::block_diagram} shows the three main attributes in the \texttt{Benchmark} class: 1) the methods to be compared, 2) the performance metric, 3) the results.

\paragraph{Methods:}
In order to make the toolbox versatile, a method is considered as a generic Python function with a variable number of input arguments using the following signature:
\begin{equation*}
  \texttt{X = method(s,\; args)}
\end{equation*}
where $\texttt{s}$ is the noisy signal, and $\texttt{args}$ represents a variable number of input arguments (either positional arguments or key-word arguments) that the user can pass to this method.
This enables the user to define what the function $\texttt{method}(\dots)$ does, and set up a number of parameter combinations that will be fed to the method using a variable number of arguments. 
The only limitation imposed by the toolbox is on the output parameter $\texttt{X}$, which depends on the task. 
In the case of the denoising methods studied in this article, $\texttt{X}$ must have the same dimensions as $\texttt{s}$. 
For detection methods, $\texttt{X}$ must be a Boolean variable, indicating whether a signal is detected in $\texttt{s}$ or not.

\paragraph{Performance metrics:}
Performance metrics are computed by generic functions that receive a vector array with the original signal $\texttt{x}$, the output of a method $\texttt{X}$, and a variable number of input parameters $\texttt{args}$:
\begin{equation*}
  \texttt{ y = performance\_metric(x,\; X,\; args)},
\end{equation*}
where $\texttt{y}\in\mathbb{R}$. 
As before, the variable number of input parameters is used to pass further information to this function that might be needed to compute the performance metric, such as the noise realizations used to contaminate $\verb|x|$.
This provides users with flexibility to implement their own performance functions.

\paragraph{Results:} The main final product of a benchmark created with MCSM-Benchs is a two-dimensional data array (technically, a Pandas \emph{DataFrame} \cite{pandas1,pandas2}) where all the results are stored to ease the analysis using specific libraries, or to be shared with other users.

\subsubsection{The \texttt{ResultsInterpreter} class}
Analyzing the data generated by the benchmarking process can be a cumbersome process. 
For this reason, a class that can generate a series of tables, figures and files, from the output of the performance metrics, is provided.
This helps in displaying the results in a similar format regardless of the user.
It is also particularly useful for those who are less familiar with Python.
As an example, the results that will be shown later in Sec. \ref{sec:results} were generated using this component of the toolbox as well as the interactive plots shown in the public repository.

\subsection{Other functionalities}
Some extra functionalities of MCSM-Benchs are summarized below that can be of interest for users:
\begin{itemize}
  \setlength\itemsep{0.1em}
    \item Methods can be run in parallel when appropriate hardware is available, reducing the benchmarking time. This can be indicated by the user as an input parameter, as well as the desired number of processes.
    \item The execution time of each method can also be computed, allowing users to take it into account in their comparisons.
    \item MCSM-Benchs provides a custom \textsc{matlab} interface that can be used to call methods implemented in \textsc{matlab} directly from the Python-based benchmark. Some restrictions on \textsc{matlab}'s version may apply in this case.
    \item Interactive figures can be easily obtained from the \verb|ResultsInterpreter| class.
\end{itemize}

\subsection{Collaborative benchmarks}
The main purpose of MCSM-Benchs is to generate benchmarks of methods based on multi-component signals that can be \emph{collaborative}, growing with the support of the community in order to be representative and useful.
One step towards this direction is the use of open-source code.
MCSM-Benchs can be downloaded from a public repository \footnote{\url{https://github.com/jmiramont/mcsm-benchs}}, where the documentation and the many examples given there can help the user to understand how to use it.

Another step towards a more general adoption of this type of practice is to make the benchmarks generated using MCSM-Benchs available for the community, using for instance public repositories of code.
These are designed to foster collaboration, by allowing other users to easily add new methods and raise issues to improve.
The typical directory tree of a benchmark created with MCSM-Benchs is similar to the following one:
\begin{center}
\begin{verbatim}
a-benchmark-repository
└── src
└    └── methods
└    └── utilities
└── config.yaml
└── run_this_benchmark.py
\end{verbatim}
\end{center}
In order to add a new method, the user just needs to add a Python file \emph{representing} the method in the folder \texttt{methods} (even for \textsc{matlab}-implemented approaches), following the instructions given in the repository. 
Template files and step-by-step guidance are provided to make this procedure more straightforward.
The benchmark can then be run using the script \verb|run_this_benchmark.py| that automatically discovers all methods in the repository and launches the benchmark with the parameters given in the \verb|config.yaml| file.
When new methods are added to a preexisting benchmark, the simulations can be run either by applying only the newly added approaches, extending previous results, or from scratch, with the purpose of reproducing the whole set of simulations.

\section{Benchmarks} \label{sec:methods}
This section describes three benchmarks that illustrate the use of MCSM-Benchs. 
The signal detection and signal denoising tasks of the benchmarks shown later in this section, are defined first. 
Moreover, this section also introduces mathematical notation and key notions, like the the short-time Fourier transform (STFT) and the spectrogram, in order to set up the context before describing the methods involved in each benchmark.
Finally, each benchmark is described, stating its goal and outlining the competing methods.
A summary of the benchmarks and the methods involved in each one can be seen in Table \ref{tab:summary}.

\begin{table*}
    \centering
        \caption{Summary of the benchmarks described in this article.} \label{tab:summary}
   \resizebox{\textwidth}{!}{%
    \begin{tabular}{|c|c|c|c|c|} \hline
         Benchmark   & Task         & Methods   & Performance Metric(s) & Signals         \\ \hline
         Benchmark 1 & Detection    & Monte Carlo Envelope test, Global MAD tests, Global Rank tests,           & Detection Power    & Synthetic       \\ \hline
         Benchmark 2 & Denoising    & STFT thresholding, Contour-based basins of attraction, Synchrosqueezing + RD     & QRF                & Synthetic       \\
                     &              & Pseudo-bayesian RD, Empty Space, Delaunay Triangulation, Spectrogram Zeros Classification &                    &                  \\ \hline         
         Benchmark 3 & Denoising    &  Hard Thresholding, Delaunay Triangulation        & QRF and APS        & Synthetic       \\ \hline
    \end{tabular}}
\end{table*}
\subsection{The tasks: signal detection and denoising}

Let $x = s + \xi$ be a continuous real- or complex-valued signal of $t\in\mathbb{R}$, where $t\mapsto s(t) \in L^2(\mathbb{R})$ is the noiseless signal, and $\xi$ is the added noise. 
What one means by \mbox{\emph{signal}} and \emph{noise} depends on the application, but usually it can be considered that the signal is a quantity of interest that exhibits some \emph{organization} and, in contrast, noise is usually regarded as a \emph{random} fluctuation including all the other influences on the observed phenomena that are without interest \cite{flandrin2018explorations}.
Throughout this work, $\xi$ will be real white Gaussian with zero mean and  variance $\sigma^2$, see e.g. \cite[Chapter 2.1]{HOUZ10} for a mathematical definition. 

Then, the Signal-to-Noise Ratio (SNR) between $x$ (signal) and $\xi$ (noise) will be defined by
\begin{equation}
 \operatorname{SNR}(s,\xi) = 10 \log_{10} 
 \left(
 \frac{\norm{s}_{2}^{2}}{
 \mathbb{E}
 \left[
 \norm{\xi}_{2}^{2}
 \right]
 }
 \right)
 \;\text{(dB)} \label{e:snr}
\end{equation}
where $\norm{\cdot}_{2}$ is the usual 2-norm, and $\norm{s}_{2}^{2}$ is the power of the signal.
The tasks of detecting a signal in noise and denoising this signal are considered.
\subsubsection{Detection} 
\label{sec:detection}
This task aims at discovering the presence of a signal $t\mapsto s(t)$ immersed in noise \cite{kay2006fundamentals, van2004detection}, and is most often formalized as a hypothesis test \cite{whalen2013detection,flandrin1998time, bardenet2018zeros}.
Two hypotheses are considered:
\begin{equation} \label{e:test}
 \left\lbrace \begin{array}{l}
               H_{0}: x = \xi \; \\
               H_{1}: x = s + \xi
              \end{array}\right. .
\end{equation}
The significance of the test, given by $\alpha$, is the probability to reject $H_0$ under $H_0$.
In contrast, the \emph{detection power} is then the probability, under $H_1$, to (correctly) reject $H_0$.
Because the tests are designed to have a prescribed value of $\alpha$, our main interest will lie in estimating the detection power, and compare it among different tests.

\subsubsection{Denoising}
The goal of a denoising method is, for a given noisy signal $x$, to approximate $s$, \emph{as well as possible}, by some $t\mapsto \tilde s(t)$, according to some performance metric such as the widely used \emph{quality reconstruction factor} (QRF)\cite{meignen2016adaptive}
\begin{equation}
 \operatorname{QRF} = 10 \log_{10}\left( \frac{\norm{s}^2_{2}}{\norm{s-\tilde{s}}^{2}_{2}} \right)\; \text{(dB)} 
\end{equation}
which can be seen as the SNR of the denoised signal, obtained by approximating the residual noise as the difference between $\tilde{s}(t)$ and the true signal $s(t)$.

\subsection{The spectrogram of multi-component signals}

\subsubsection{The STFT}

The STFT can be considered as a time-localized version of the Fourier transform. It is a fundamental tool for TF signal analysis. Given a signal $x\in L^{1}\left( \mathbb{R} \right) \cap L^{2}\left( \mathbb{R} \right)$, its STFT is defined as 
\begin{equation}\label{e:stft}
V_{x}^{g}(t,\nu) = \int_{-\infty}^{+\infty} x(u)g(u-t)^{*}e^{-i2\pi \nu u} du,
\end{equation}
where $g(t)\in  L^{2}\left( \mathbb{R} \right)$ is the analysis window.
In practice, however, one computes a discrete approximation of Eq. \eqref{e:stft} for a discrete signal $x[n]$ of period $N$, given by
\begin{equation}\label{e:stft_dis}
  V_{x}^{g}[n,k] = \sum\limits_{\ell = 0}^{N-1} x[\ell]g[\ell-n] \;e^{-i \frac{2\pi \ell k }{K} },
\end{equation}
where $n=0,1,\dots,N-1$ and $k=0,1,\dots,K$ are, respectively, the discrete time and frequency indexes, and $g[n]$ a unit-energy Gaussian window of width $T$
\begin{equation} \label{e:window}
g[n] = \frac{2^{1/4}}{\sqrt{T}} e^{- \pi (\frac{n}{T})^2}.
\end{equation}

\subsubsection{Multi-component signals}

Numerical experiments will deal with multi-component, amplitude-modulated, frequency-modulated (AM-FM) signals, of the form
\begin{equation}
  s[n] = \sum_{j=1}^{J} s_{j}[n], \text{ with } s_{j}[n] = a_{j}[n]e^{i \phi_{j}[n]},
\end{equation}
where $J$ is the number of signal components, $s_{j}[n]$ is an AM-FM component or \emph{mode}, and $a_{j}[n]$ and $\phi_{j}[n]$ are the instantaneous amplitude and phase functions of the $j$-th component, respectively.
Each mode can be associated with a \emph{ridge} in the spectrogram that describes the trajectory of the instantaneous frequency given by $\frac{1}{2\pi}\phi^{\prime}(nT_{s})$ where $\phi^{\prime}(t)$ is the continuous-time counterpart of the phase of the $j$-$th$ mode and $T_{s}$ is the sampling period.
Detecting the ridges is the cornerstone of the methods based on the largest values of the spectrogram, some of which are outlined in the following section.
\begin{figure}
  \centering

  \begin{tabular}{cccc}
\multicolumn{4}{c}{\includegraphics[width=0.48\textwidth]{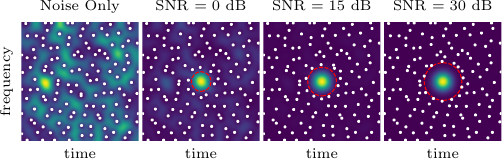}} \\
  \end{tabular}
  
  \caption{Spectrograms of noise and of three mixtures of a first-order Hermite function and noise, with different SNRs. White dots indicate the position of the zeros of the spectrogram. The red dashed line approximately marks the border of the signal domain $\mathcal{D}_{s}$.} \label{fig::shrinking_domains}
\end{figure}
\subsubsection{Zeros versus large values}

The spectrogram of $x(t)$ (resp. $x[n]$) is then defined as the squared modulus of $V_{x}^{g}(t,\nu)$ (resp. $V_{x}^{g}[n,k]$), and can be thought of as a TF distribution of the energy of the signal. 
One can then consider that the signal is stronger where its energy is more concentrated in the TF plane \cite{cohen1966generalized, auger1995improving, flandrin1998time}.

In contrast with the paradigm of informative large values of the spectrogram, it has been shown that its zeros are relevant points, intimately related to the structure of the components in the TF plane \cite{flandrin2015time}. 
The zeros of the spectrogram of white noise seem to be very regularly distributed in the time-frequency (TF) plane (as shown in Fig. \ref{fig::shrinking_domains}a).
Indeed, it has been shown that their distribution corresponds to that of the zeros of the planar Gaussian analytic function \cite{flandrin2015time, bardenet2018zeros}, a well-studied point process exhibiting various degrees of organization \cite{hough2009zeros}. 
In the case of a noisy signal, the spectrogram zeros tend to surround the signal domain $\mathcal{D}_{s}$, i.e. the region of the TF plane with more information on the signal, producing large regions without zeros (or \emph{empty} regions) (as shown in Figs. \ref{fig::shrinking_domains}b to \ref{fig::shrinking_domains}d) \cite{flandrin2015time, flandrin2018explorations}.
This fact can be exploited to detect the presence of a signal in a mixture of signal and noise by means of Monte Carlo tests commonly used in spatial statistics \cite{baddeley2014tests, bardenet2018zeros}, and to estimate $\mathcal{D}_{s}$ for denoising \cite{flandrin2015time,bardenet2018zeros,ghosh2022signal,pascal2022covariant}.

\subsection{Benchmark 1. Signal detection tests based on the spectrogram zeros} \label{sec:tests}
Considering the zeros of the spectrogram as a spatial point process $\mathcal X$ in $\mathbb{C}\approx\mathbb{R}^2$ \cite{hough2009zeros, baddeley2015spatial}, one can summarize its distribution by functional quantities like the empty space function 
\begin{equation}
  \label{e:F}
  F(r) = \mathbb{P}(\mathcal{X} \cap B(0,r)\neq \emptyset), \quad r>0,
\end{equation}
where $\mathbb{P}(\mathcal{X}\cap B(0,r)\neq \emptyset)$ is the probability that an Euclidean ball $B(0,r)$ centered at $0$, with radius $r$, contains at least one point of $\mathcal{X}$.
Many such \emph{summary} functions can be defined and estimated, to investigate the behavior of spatial point processes that are invariant to rotations and/or translations \cite{baddeley2014tests,baddeley2015spatial,myllymaki2017global}.
Henceforth, let $r\mapsto S(r)$ denote a summary function, with $r\mapsto \hat{S}(r)$ an estimator of that function.

This benchmark compares the detection power of three different types of detection tests based on Monte Carlo simulations: the Monte Carlo envelope test \cite{baddeley2014tests,bardenet2018zeros}, the global maximum absolute deviation (MAD) test \cite{baddeley2014tests} and the global rank envelope test \cite{myllymaki2017global}. These are based on comparing, on one side, the realization $r\mapsto \hat{S}_{0}(r)$ of the estimator of $S$ obtained from the observed signal, with, on the other side, $m\gg 1$ realizations $\hat S_{1},\dots, \hat S_{m}$ of the same estimator obtained from independent simulations under the null model $H_{0}$, for $r$ in some interval $I = [r_{\min},r_{\max}]$.
For demonstration purposes, this paper considers either $S=F$, the empty space function in Eq. \eqref{e:F}, or its variance-stabilized version \cite{baddeley2014tests, najari2022neyman} 
\begin{equation*}
\widetilde F:r\mapsto \arcsin(\sqrt{F(r)}).
\end{equation*}

For the estimator $\hat{S}$, the ``border correction'' estimator \cite{baddeley2015spatial} was chosen in both cases.
Other commonly used functional statistics, like Ripley's $K$, were previously studied in \cite{bardenet2018zeros}. The following describes the particularities of each of the mentioned tests.

\subsubsection{Monte Carlo envelope tests} 
For a Monte Carlo \emph{envelope} test, one first chooses the significance level $\alpha$, and then fix an integer $k\leq m$ such that $\alpha = k/(m+1)$.
Next, $m$ independent realizations of the white noise spectrogram are generated, and one uses their $m$ sets of zeros to evaluate the corresponding $\hat{S}_{1}, \dots, \hat{S}_{m}$.
Similarly, denote the functional statistic evaluated on the data by $\hat{S}_{0}$. 
To derive real numbers from these estimated functions, one natural option is to define
\begin{equation}
t_{j} = \left\Vert(\hat S_{j}(\cdot) - \bar S(\cdot)) \mathbbm{1}_{[r_{\min},r_{\text{MC}}]} \right\Vert_{p},\; j=0,\dots,m,
\label{e:test_statistic}
\end{equation} 
where $0<r_{\text{MC}}<r_{max}$ and $p>0$ are parameters of the test, $r\mapsto \bar S(r)$ is the pointwise average of $\hat S_{0},\dots,\hat S_{m}$ (which \emph{includes} the observation being tested), and $r_{\text{MC}}$ is selected \emph{a priori}, generally based on the knowledge of the problem in question \cite{baddeley2014tests,myllymaki2017global}.

Now, for fixed values of $p$, $r_{\min}=0$ and $r_{\text{MC}}>r_{\min}$ in Eq. \eqref{e:test_statistic}, the values $t_1,\dots,t_m$ are sorted. 
Intuitively, a large $t_{0}$ among the values obtained by sampling under $H_0$ is suspicious, hence the null hypothesis is rejected when $t_{0}\geq t_{(k)}$, where $t_{(k)}$ is the $k$-th largest $t_{j}$.
A symmetry argument \cite{baddeley2014tests} guarantees that this detection test has the prescribed significance level $\alpha$.

\subsubsection{Global MAD test} 
This test \cite{baddeley2014tests} is based on taking the maximum absolute deviation between $S_{0}$ and a theoretical value $S_{\text{theo}}$ on the entire interval $I$ or, equivalently use $r_{\text{MC}} = r_{\max}$ and the supremum norm in Eq. \eqref{e:test_statistic}.

This amounts to consider whether $\hat{S}_{0}(r)$ is greater than $t_{(k)}$ for at least one value of $r \in I$.
The test is commonly called \emph{global} for its avoidance to fix a subjective value for $r_{\text{MC}} \in I$.

\subsubsection{Global rank envelope test} 
This test is based on assigning a rank $\rho_j$ to each estimated function $r\mapsto \hat S_j(r)$, defined as \cite{myllymaki2017global}
\begin{equation}
  \rho_{j} = \max \left\lbrace k: S_{\text{low}}^{(k)}(r) \leq \hat S_{j}(r) \leq S_{\text{upp}}^{(k)}(r),  \forall r\in I \right\rbrace,
\end{equation}
where $S_{\text{low}}^{(k)}$ and $S_{\text{upp}}^{(k)}$ are the $k$-th lower and upper envelopes, defined for $r>0$ as
\begin{equation}
  S_{\text{low}}^{(k)}(r) = \!\!\underset{j=0,\dots,m}{\min^k} \, \hat S_{j}(r) 
  \text{ and }
  S_{\text{upp}}^{(k)}(r) = \!\!\underset{j=0,\dots,m}{\max^k} \, \hat S_{j}(r),
\end{equation}
where $\min\limits^k$ ($\max\limits^k$) is the $k$-th smallest (largest) value.
One can consider $\rho_{j}$ as a measure of how ``extreme'' is $\hat{S}_{j}$ when compared with the ensemble $\hat{S}_{1},\dots,\hat{S}_{m}$ \cite{myllymaki2017global}.
If $\hat S_{0}$ is too extreme when compared to the simulated $\hat S_{j}$, one can suspect that it does not come from the null model $H_{0}$.
A range of $p$-values $[p_{-}, p_{+}]$ can be computed \cite{myllymaki2017global}, and the authors respectively define a \emph{liberal} and a \emph{conservative} test by rejecting whenever $p_-$, respectively $p_+$, is smaller than $\alpha$.
As a rule of thumb, a number $m=2499$ of simulations is recommended to make these two tests essentially indistinguishable when the significance level is $\alpha=0.05$ \cite{myllymaki2017global}. 

\subsection{Benchmark 2. Contrasting Paradigms: Large Values vs. Zeros of the Spectrogram}
\label{sec:benchmark_2}

The length of the signals used in this benchmark was set to $N=1024$ time samples, and the results were computed as the average over $100$ repetitions of the simulations with different noise realizations.
Next sections introduce each of the approaches compared in this benchmark.

\subsubsection{Short-time Fourier transform thresholding} \label{sec:method_thr}
The basic premise of this technique is to remove the coefficients of the STFT according to some thresholding function \cite{gao1998wavelet, donoho1994ideal, mallat2008tour}, in order to obtain a filtered version of the STFT given by $\widetilde{V}_{x}^{g}$.

To compute the threshold, the following estimation of the standard deviation of a (real white) Gaussian noise is considered \cite{donoho1994ideal,mallat2008tour}: 
 \begin{equation} \label{e:ht_estimator}
  \hat{\sigma} = \frac{\sqrt{2}}{0.6745}\operatorname{median}\left\lbrace | \real \{ V^{g}_{x}\} | \right\rbrace.
 \end{equation}
The threshold is then defined as $\lambda = c\; \hat{\sigma}$, where $\hat{\sigma}$ is estimated using Eq. \eqref{e:ht_estimator} and $c$ varies with the thresholding function.
Two different thresholding approaches\cite{donoho1995noising, gao1998wavelet} are explored (dropping the variables $[n,k]$ here for a more succinct notation):
\begin{enumerate}
  \item Hard Thresholding (T-Hard): 
  \begin{equation*}
    \widetilde{V}_{x}^{g}=\left\lbrace
                            \begin{array}{ll}
                              0 &\text{ if } |V_{x}^{g}| \leq \lambda \\
                              V_{x}^{g} & \text{ if } |V_{x}^{g}| > \lambda 
                            \end{array}\right.,
  \end{equation*}
where $\lambda = 3.0\; \hat{\sigma}$.

  \item Soft Thresholding (T-Soft, with \emph{Garrote} function \cite{gao1998wavelet}): 
  \begin{equation*}
    \widetilde{V}_{x}^{g}=\left\lbrace
                            \begin{array}{ll}
                              0 &\text{ if } |V_{x}^{g}| \leq \lambda \\
                              (1-\frac{\lambda^2}{|V_{x}^{g}|^2})V_{x}^{g} & \text{ if } |V_{x}^{g}| > \lambda \\
                            \end{array}\right.,                            
  \end{equation*}
  where $\lambda = 2.0\; \hat{\sigma}$.
  \end{enumerate}
  The value of $c$ for each method was empirically chosen with the aim of maximizing the performance of the thresholding methods for most of the signals used in the benchmark.
  The signal can then be estimated by inverting $\widetilde{V}_{x}^{g}$ as
  \begin{equation} \label{eq::invert_th}
    \tilde{s}[n] =  \frac{1}{Kg(0)} \sum\limits_{k=0}^{K-1} \widetilde{V}^{g}_{x} [n,k]\; e^{i \frac{2\pi n k}{K} }.
  \end{equation}

  \subsubsection{Contour-based basins of attractions} \label{sec:method_contours}
  The ridges generated by the signal components are known to be \emph{attractors} of the reassignment vector \emph{field} \cite{auger1995improving, chassande1997differential,meignen2016adaptive,flandrin2018explorations}
  \begin{equation}
    RV: [n,k] \mapsto \left( \hat{\tau}[n,k]-n,\; \hat{\nu}[n,k]-k \right)
  \end{equation}
  with 
  \begin{equation}
    \hat{\tau}[n,k]= n+ \real\left\lbrace \frac{V^{ng}_{x}[n,k]}{V^{g}_{x}[n,k]} \right\rbrace, 
  \end{equation}
 \begin{equation}\label{eq:rm_operators}
    \hat{\nu}[n,k]= k - \frac{K}{2\pi}\imaginary\left\lbrace \frac{V^{g^{\prime}}_{x}[n,k]}{V^{g}_{x}[n,k]} \right\rbrace,
  \end{equation}
where $V^{ng}_{x}$ and $V^{g^{\prime}}_{x}$ are the discrete STFTs computed with windows $n\mapsto ng[n]$ and $n\mapsto g^{\prime}[n]$, respectively. 

The ridges can then be detected by interpreting $RV[n,k]$ as a vector and computing the projection
  \begin{equation} 
    v[n,k] = \left\langle RV[n,k], \gamma_{\theta}  \right\rangle,
  \end{equation}
  where $\gamma_\theta$ is a unit vector, and $\theta$ is the projection angle given by $\theta[n,k] = \angle{RV[n,k]}\operatorname{mod} \pi$, where $\angle{RV[n,k]}$ is the argument of the complex-valued $RV[n,k]$.
  Then, the contours given by $v[n,k]=0$ correspond to the ridges and the valleys of the spectrogram \cite{lim2012sparse,meignen2016adaptive}.
  The implementation proposed in \cite{pham2017adaptive} is used, where the projection angle at any given TF point in the discrete grid is estimated as the most frequent value of $\theta[n,k]$ in a square neighborhood of the point. 
  Although computationally costly, such an implementation is more robust to noise, and can even extract impulsive transients \cite{pham2017adaptive}.

  Because the zeros of the spectrogram are \emph{repellers} of $RV$ \cite{chassande1997differential,meignen2016adaptive}, the contours can be segmented into a set of curves by using the zeros as beginning and end points. 
  The TF support of each signal component can then be estimated by considering the contours corresponding to ridges, and estimating their associated \emph{basins of attraction}, formed by the points in the TF plane that would be relocated to a given ridge by the reassignment method \cite{auger1995improving, chassande1997differential}.
  The contour-based method for signal estimation consists in computing the level curves $v[n,k]=0$, segmenting them and determining, for each ridge, its corresponding basin of attraction, keeping the $J$ more energetic ones associated with the signal components $s_{j},j=1,\dots,J$.
  Note that the number of components needs to be known in advance.
  Once the ridges and their basins of attractions are identified, the individual components can be synthesized as
  \begin{equation} \label{eq::invert_basins}
    \tilde{s}_{j}[n] =  \frac{1}{Kg(0)} \sum\limits_{k=0}^{K-1} \widetilde{V}^{g}_{x} [n,k] \mathbbm{1}_{\mathcal{B}_{s_{j}}}[n,k] \; e^{i \frac{2\pi n k}{K} },
  \end{equation}
  where $\mathbbm{1}_{\mathcal{B}_{s_{j}}}$ is a $1$/$0$ mask corresponding to the basin of attraction $\mathcal{B}_{s_{j}}$ associated to the estimated component $\tilde{s}_{j}$.
  Henceforth, this approach is called the \emph{Contours} method.

\subsubsection{Synchrosqueezing transform and ridge detection} \label{sec:method_brevdo}
Synchrosqueezing (SST) is a post-processing transformation that produces a sharper TF representation while being invertible \cite{daubechies2017nonlinear,daubechies2011synchrosqueezed}.
It \emph{vertically} reassigns the STFT coefficients according to $\hat{\nu}[n,k]$ defined in Eq. \eqref{eq:rm_operators}, 
\begin{equation}
  \mathcal{T}_{x}[n,k] = \sum\limits_{q: \left| k-\hat{\nu}[n,q]\right|\leq1/2} V^{g}_{x}[n,q]\; e^{i\frac{2\pi q n}{K}}.
\end{equation}

When combined with a ridge detection algorithm \cite{thakur2013synchrosqueezing, carmona1999multiridge} it allows extracting every individual signal components as
\begin{equation} \label{eq:invert_modes}
  \tilde{s}_{j}[n] = \frac{1}{Kg(0)}\sum\limits_{q\in\left[\Omega_{j}[n]-\epsilon/2,\Omega_{j}[n]+\epsilon/2\right]} \mathcal{T}_{x}[n,q], 
\end{equation}
where $\Omega_{j}$ is the estimated ridge corresponding to the $j\text{-th}$ mode, and $\epsilon>0$ is the width of a strip around $\Omega_{j}$, usually fitted to the support of the Fourier transform of the analysis window defined in Eq. \eqref{e:window}.
Throughout the following sections, this method is referred as the \emph{SST+RD} method. 

\subsubsection{Pseudo-Bayesian ridge detection} \label{sec:method_pb}
This method is based on detecting ridges in the discretized spectrogram $[n,k] \mapsto |V^{g}_{x}[n,k]|^{2}$. 
The algorithm \cite{legros2021novel,legros2022pb} is an elaborate construction based on the following principle: a ridge is a sequence of points in different columns of the spectrogram, each of these points being the mean of a Gaussian fitted to its column. 
Robust fitting is carried out by minimizing an $\alpha$-$\beta$ cross-entropy, with parameters $\alpha,\beta \in \mathbb{R}$, $\alpha+\beta\neq 0$, $\alpha\neq 0$, $\beta\neq 0$ (see \cite{legros2022pb} for further details).
We explore two combinations of parameters for this method $\alpha=\beta=0.4$ and $\alpha=0.2$, $\beta=0.4$, as described in \cite{legros2022pb}.

\subsubsection{Aggregation of \emph{empty} TF regions without zeros} \label{sec:method_es}
As explained before, the homogeneous distribution of the zeros of the spectrogram of noise is \emph{locally} disrupted by the presence of a signal component in the TF plane \cite{flandrin2015time, bardenet2018zeros} and larger-than-expected regions without zeros corresponding to the support of the signal the plane. 
Consequently, the detection of these zones is akin to estimating $\mathcal{D}_{s}$.
This is precisely the rationale of the method proposed in \cite{bardenet2018zeros}, called here the \emph{Empty Space} (ES) method, where $\mathcal{D}_{s}$ is formed by aggregating Euclidean balls of radius $r_{0}$ that do not contain any zero \cite{bardenet2018zeros}.

The Empty Space method can be summarized in the following steps:
\begin{enumerate}
  \item Compute the spectrogram of $x$, using $T=\sqrt{K}$ in Eq.  \eqref{e:window}.
  \item Detect the zeros of the spectrogram as local minima of $[n,k] \mapsto |V^{g}_{x}[n,k]|^2$.
  \item Find all Euclidean balls of radius $r_{0}$ without zeros in the (discrete) TF plane.
  \item Estimate $\mathcal{D}_{s}$ as the union of those balls.
  \item Reconstruct the denoised signal using
  \begin{equation} \label{eq:invert_sd}
    \tilde{s}[n] =  \frac{1}{K g(0)} \sum \limits_{k = 0}^{K-1} V^{g}_{x} [n,k]  \mathbbm{1}_{\mathcal{D}_{s}}[n,k] \;e^{i \frac{2\pi n k}{K} },
  \end{equation}
where ${1}_{\mathcal{D}_{s}}[n,k]$ is a 1/0 mask aiming to extract the estimated signal domain.
\end{enumerate}

\begin{figure}
  \centering
   \setlength{\tabcolsep}{-2pt}
   \renewcommand{\arraystretch}{0.1}
    \begin{tabular}{c}
      \multicolumn{1}{c}{\includegraphics{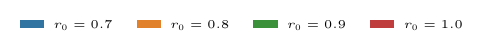}} \\
      \includegraphics{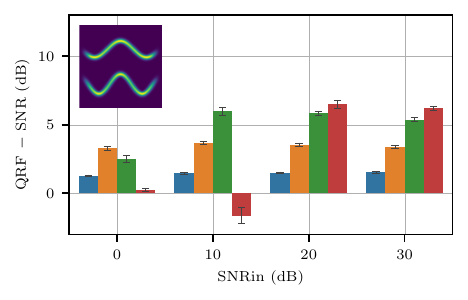}  
    \end{tabular}
    \caption{Difference between the Quality Reconstruction Factor (QRF, in dB) and the input SNR computed after filtering a signal using the Empty Space method described in Sec. \ref{sec:benchmark_2} and in \cite{bardenet2018zeros}. Each bar shows the mean for $100$ repetitions of the experiment for a fixed $r_{0}$. The errobars indicate the $95$\% confidence interval with Bonferroni correction.} \label{fig::params_es}
  \end{figure}

As shown in Fig. \ref{fig::shrinking_domains}, $\mathcal{D}_{s}$ \emph{shrinks} when the SNR of the signal decreases and, consequently, so does $r_{0}$.
This effect is also illustrated in Fig. \ref{fig::params_es}, which shows the difference between the QRF and input SNR (dB), which can be interpreted as the improvement (in dB) in the SNR of the signal after filtering it with the Empty Space method. 
The spectrogram of the signal used in this case is also shown in Fig. \ref{fig::params_es}.
By observing the bars corresponding to the fixed values of $r_0 \in \{0.7, 0.8, 0.9, 1.0\}$ in Fig. \ref{fig::params_es}, one can to see that the appropriate value of $r_{0}$ decreases with the SNR, as expected.
Since the SNR is not known in practice, one has to manually tune $r_0$.
In order to improve the method, a strategy is settled in the following to automatically estimate $r_0$.

    \begin{figure}
    \centering
    \setlength{\tabcolsep}{-2pt}
    \renewcommand{\arraystretch}{0.1}
    \begin{tabular}{c}
      \includegraphics{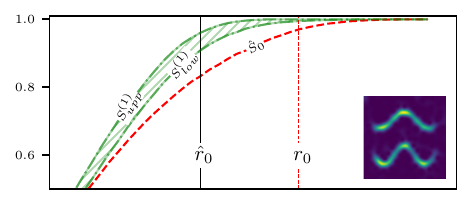} \\
      (a) \\
      \includegraphics{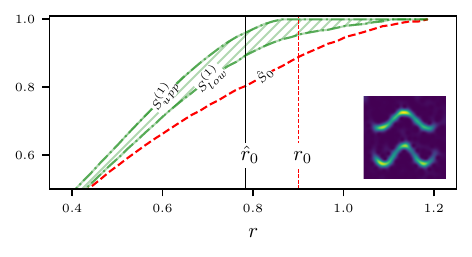} \\
      (b)
    \end{tabular}
    \caption{(a) Upper ($S^{(1)}_{\text{upp}}$) and lower ($S^{(1)}_{\text{low}}$) envelopes, in green, dot-dashed lines, the empirical functional statistic $\hat{S}_{0}$, in red, dashed line, and the value of $r_{0}$ estimated by Algorithm \ref{alg::adapt_r0}, indicated by a vertical, continuous line, for $S=F$. (b) Idem (a), but for $S = \widetilde{F}$. The spectrogram shown of the used signal is shown. The vertical, red, dashed line indicates the optimal $r_{0}$, according to Fig. \ref{fig::params_es}.} \label{fig::envelopes}
  \end{figure} 

Ideally, one would like to select $r_{0}$ as the value of $r$ that causes the rejection of the null hypothesis in one of the tests described in Sec. \ref{sec:tests} \cite{bardenet2018zeros}.
Such value of $r$ is termed the \emph{scale of interaction}.
The parameter $r_{0}$ will be estimated as the value of $r$ corresponding to the maximum difference between $\hat{S}_{0}(r)$ and the lowest envelope $S_{\text{low}}^{(1)}(r)$ for $r\in I$, provided that $H_0$ was rejected using one of the tests described above.
Formally, $r_{0}$ is then estimated as 
\begin{equation}
  \hat{r}_{0} = \argmax_{r\in I} |S_{\text{low}}^{(1)}(r) - \hat{S}_{0}(r)|.
\end{equation}

Therefore a detection test and a summary function $S(r)$ need to be chosen.
The test uses the Global Rank Envelope test described before with $I=[0.65;1.05]$. 
This is because, as will be shown later in Sec. \ref{sec:results:b1}, it is one of the tests with the best performance.
As for $S$, we study how good this estimation is when \mbox{$S=F$}, as proposed in \cite{bardenet2018zeros}, and when $S = \widetilde{F}$.
 
Fig. \ref{fig::envelopes}a shows that using $F$ leads to an underestimation of $r_{0}$ (the optimal value is indicated with a vertical red, dashed line).
Meanwhile, $\widetilde{F}(r)$ increases the difference between the lower envelope and the observed statistic for larger values of $r$, reducing the gap between the estimated $r_0$ and the optimal value.

Computing the scale of interaction is then conditioned upon the signal detection, i.e. the null hypothesis must be rejected at the end of the detection test, which is why a test with a high detection power must be used.
However, if detection fails, the value of $r_{0}$ is set to $0.8$ following the observation that this value is nearly optimal for the lowest SNRs used here for most of the signals (as shown, for example, in Fig. \ref{fig::params_es}).
The steps taken to estimate $r_{0}$ are summarized in Algorithm \ref{alg::adapt_r0}.

\begin{algorithm}[htbp]
  \begin{algorithmic}[1]
  \REQUIRE The signal $x$, the number of simulations $m$.
  \STATE Compute $\hat{S}_{0}(r),\hat{S}_{1}(r),...,\hat{S}_{m}(r), r\in [0.65;1.05]$ for $S=\widetilde{F}$.
  \STATE Compute the Global Rank Envelope Test (Sec. \ref{sec:tests}).
  \IF{$H_{0}$ is rejected}
  \STATE Compute $S_{\text{low}}^{(1)}(r),\; r\in [0.65;1.05]$.
  \STATE Estimate $r_{0}$ as 
  \begin{equation*}
    \setlength{\abovedisplayskip}{0pt}
    \setlength{\belowdisplayskip}{0pt} 
    r_{0} = \argmax_{r\in [0.65;1.05]} |\hat{S}_{\text{low}}^{(1)}(r) - \hat{S}_{0}(r)|.
  \end{equation*}
  \ELSE
  \STATE Set $r_{0}=0.8$. 
  \ENDIF
  \RETURN $r_{0}$
  \end{algorithmic}
  \caption{Adaptive estimation of $r_{0}$.} \label{alg::adapt_r0}
  \end{algorithm}

\subsubsection{Delaunay triangulation of zeros} \label{sec:method_dt}
With the purpose of identifying TF support of the signal, it was proposed in \cite{flandrin2015time} to compute a Delaunay triangulation (DT) of the zeros of the spectrogram.
Because of the displacement of zeros caused by the presence of a signal, the triangles along the signal domain are more stretched than those generated only by noise, allowing its identification by selecting the triangles with an edge longer than a given value $\ell_{\max}$.
The method can be summarized in the following steps:
\begin{enumerate}
  \item Compute the spectrogram of $x$, using $T=\sqrt{K}$ in Eq. \eqref{e:window}.
  \item Detect the zeros of the spectrogram as local minima of $[n,k] \mapsto |V^{g}_{x}[n,k]|^2$.
  \item Compute a Delaunay triangulation using the spectrogram zeros as vertices.
  \item Find all triangles with an edge longer than the threshold $\ell_{\max}$.
  \item Estimate $\mathcal{D}_{s}$ as the union of the selected triangles.
  \item Reconstruct the signal using Eq. \eqref{eq:invert_sd}.
\end{enumerate}

The circumcircles of the selected Delaunay triangles play a similar role to the empty disks described above for the Empty Space method.
This suggests that one could try to use \mbox{$\ell_{\max}\approx 2 r_{0}$} to also make this approach adaptive to the signal using Algorithm \ref{alg::adapt_r0}, since this is approximately true for the elongated triangles within the signal domain.

\subsubsection{Spectrogram zeros classification} 
This method is based on classifying the spectrogram zeros in three classes according to the nature of the components interfering to produce them: 1) Signal-Signal (SS) zeros, 2) Signal-Noise (SN) zeros and 3) Noise-Noise (NN) zeros \cite{laurent2023novel, miramont2023unsupervised}.
Those of the SS kind are assumed to be produced when to signal components are very close to each other, whereas the SN kind correspond to the zeros located in the boundary of the signal support in the TF plane.
On the opposite, NN zeros are those located in the noise-only regions of the TF plane.
This method then works in a similar matter as the DT approach, keeping the Delaunay triangles associated with the SS and SN kinds zeros, and discarding those having NN zeros as a vertex.
To classify the zeros, features are extracted from 2D histograms of zeros in the TF plane.
Such histograms are built by adding an ensemble of independent noise realizations to the signal, the variance of which is estimated using Eq. \eqref{e:ht_estimator}.
This method is referred as ``Spectrogram Zeros Classification'' (SZC).

\subsection{Benchmark 3.}
Thresholding approaches, such as those described in the previous sections, as well as the DT method, suppress noise by subtracting noise coefficients from the STFT. 
Even though thresholding can be optimal from the point of view of minimizing the reconstruction error \cite{donoho1994ideal,gao1998wavelet}, one can also consider other measures of performance of more practical meaning, such as those that quantify intelligibility or audio artifacts \cite{torcoli2018comparing, torcoli2021objective}.
It is well known  \cite{goh1998postprocessing, torcoli2019improved} that hard and soft thresholding can produce a particular kind of perceptually unpleasant artifact termed \emph{musical noise}, the origin of which lies in the isolated maxima, and sometimes short-lived ridges, that are left in the spectrogram after processing the signal.
However, the type of artifacts introduced by approaches based on the spectrogram zeros is not well documented.

With this in mind, this benchmark studies the trade-off between signal reconstruction and introduction of musical noise, after applying Hard Thresholding and the DT method, using two performance metrics.
The signal reconstruction performance is quantified as previously by the QRF.
On the other hand, MCSM-Benchs can be used to create a benchmark with a user-provided performance metric, where the musical noise content is quantified by means of the \emph{artifact-related perceptual score} (APS) \cite{emiya2011subjective}.
This metric is computed by, first, using a model of auditory perception to quantify the effect of the artifact-related components of the estimation error by a salience score.
Then, the latter is mapped to audio-quality perceptual ratings produced by human listeners, using a neural network \cite{vincent2012improved}.
APS is an audio quality metric, sensitive to artifacts introduced by a source separation method. 
Hence, the higher the value of APS, the less musical noise is present in the signal.

In order to evaluate the trade-off between signal estimation and musical noise, a synthetic signal of length $N=8196$ comprising four AM-FM components is used.
Finally, these results are completed by showing the estimation of APS for a real-world signal for the two approaches studied in this benchmark.

\section{Numerical experiments}  \label{sec:results}
 This section reports the results corresponding to each benchmark. 
 
 \begin{figure*}[htbp]
  \centering
  \begin{tabular}{c}
    \includegraphics[width = \textwidth]{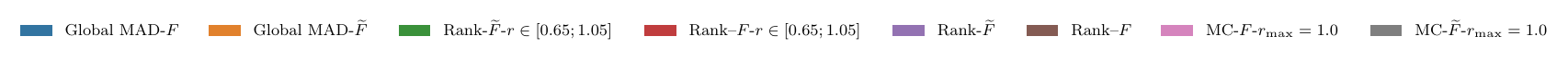} \\ 
    \includegraphics[width = \textwidth]{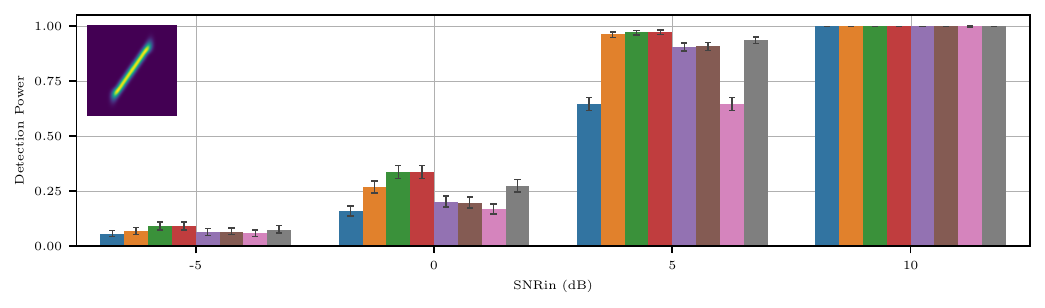}     
  \end{tabular}
    \caption{Detection power (i.e. the probability of correctly rejecting $H_{0}$) for several SNRs. Each bar shows the mean for $2000$ repetitions of the experiment. The error bars denote the 95\% Clopper-Pearson confidence intervals with Bonferroni correction.} \label{fig::power_test}
  \end{figure*}
  
  \begin{figure*}[htbp]
\centering
   \resizebox{\textwidth}{!}{%
\renewcommand{\arraystretch}{0}
\setlength{\tabcolsep}{1pt}
  \begin{tabular}{cc}
    \multicolumn{2}{c}{\includegraphics{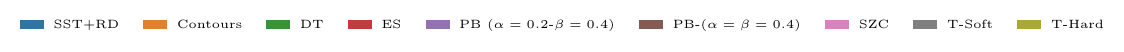}}\\
    \includegraphics{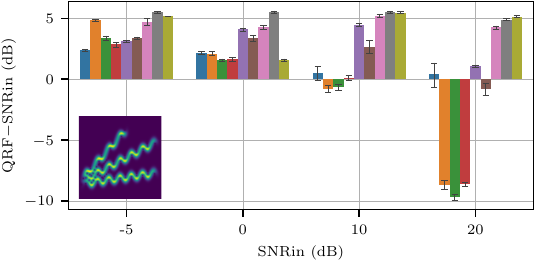} & \includegraphics{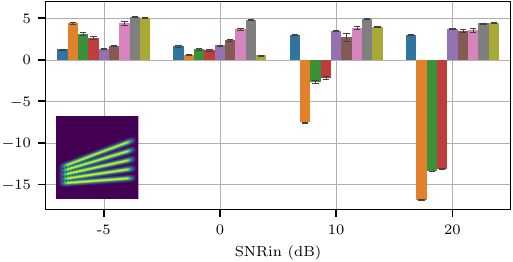} \\
    (a) & (b) \\
    \includegraphics{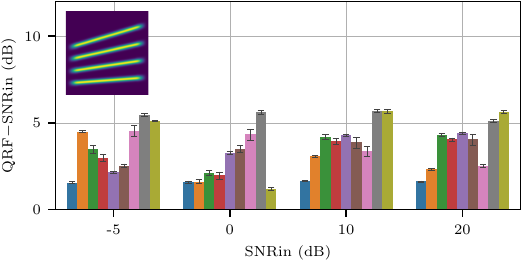} & \includegraphics{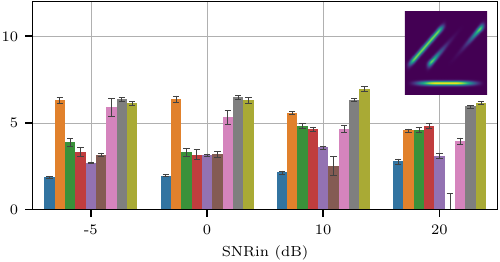} \\
    (c) & (d) \\
    \includegraphics{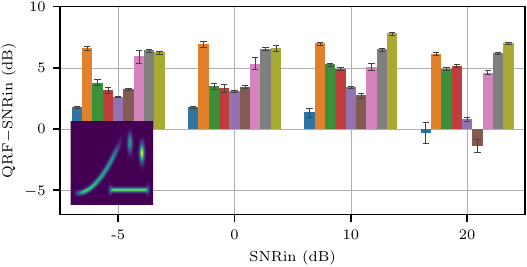} & \includegraphics{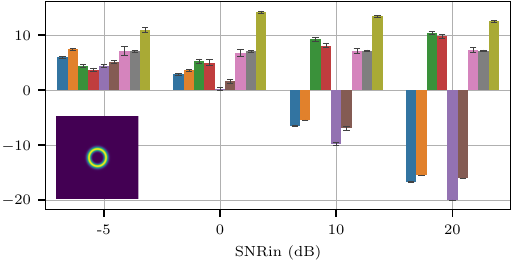} \\
    (e) & (f) \\
  \end{tabular}}
  \caption{Results of Benchmark 2. Contrasting Different Paradigms: zeros vs. large values of the spectrogram. Each bar represents the mean over $100$ repetitions of the signal estimation. The errobars indicate the $95$\% confidence interval with Bonferroni correction. 
  } \label{fig::results_denoising}
\end{figure*}

\subsection{Benchmark 1. Signal detection methods based on the spectrogram zeros} \label{sec:results:b1}

Figure \ref{fig::power_test} summarizes the obtained results in this benchmark. 
It can be seen that, for SNRs greater than $10$ dB, the performance of all tests is satisfactory.

On the other hand, for lower SNRs, the global rank tests \cite{myllymaki2017global} are more effective than the other two types of Monte Carlo tests.
A total of $m=2499$ simulations were used in all the tests, coinciding with the suggested amount for the rank envelope tests \cite{myllymaki2017global}.

Using the variance-stabilized statistic $\widetilde{F}$ increases the detection power in the cases of MAD and pointwise tests, whereas for Rank global tests, no difference is found between using $F$ or $\widetilde{F}$. 
However, reducing the considered interval $I$ for $r$ does lead to an increment in the performance for those tests.

The global MAD tests and the Monte Carlo envelope methods (MC-$F$-$r_{\max}$ and MC-$\widetilde{F}$-$r_{\max}$ in Fig. \ref{fig::power_test}) seem to have a comparable detection powers. 
Nevertheless, one should remember that the latter method require an estimation of the scale of interaction $r_{MC}$, which is not needed for the global tests.
Hence, these results seem to indicate that global tests are a better option for signal detection, requiring fewer parameters than their pointwise counterparts.

\begin{figure}

\centering
  \setlength{\tabcolsep}{1pt}
  \renewcommand{\arraystretch}{0.1}
  \begin{tabular}{ccc}
  \includegraphics{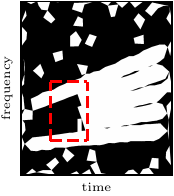} &
  \includegraphics{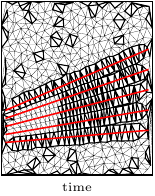} &
  \includegraphics{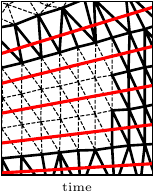} \\
  (a) & (b) & (c)     
  \end{tabular}

    \caption{(a) Extraction mask obtained by applying the Delaunay Triangulation method to the signal in Fig. \ref{fig::results_denoising}a with an SNR of $30$ dB. (b) The Delaunay triangles obtained by the method. Bold line indicates the selected triangles included in the extraction mask. Red lines indicate the instantaneous frequency of each signal component. (c) Detail of the red square shown in (a).}\label{fig::error_dt}

\end{figure}

\subsection{Benchmark 2. Contrasting paradigms: zeros vs. large values of the spectrogram}
This section reports the results of the second benchmark, where methods related to two paradigms for TF signal domain estimation were compared, using different signals.

\subsubsection{Zero-based methods}
Fig. \ref{fig::results_denoising}a shows that the two zero-based methods perform poorly for the signal consisting in three chirps with sinusoidal frequency modulations, the spectrogram of which is shown in the same figure, in particular for high SNR.
This is also the case for the signal used in Fig. \ref{fig::results_denoising}b.
Fig. \ref{fig::results_denoising}c, shows the results for a similar signal.
In this case, however, the performance is better, and all the methods manage to improve the SNR of the recovered signal.
The only difference between the signals used in Figs. \ref{fig::results_denoising}b and \ref{fig::results_denoising}c, is that the components of the latter are more spread out.

A detailed inquiry on this difference reveals that the strong interference between components occurring in the leftmost part of the signals used in Figs. \ref{fig::results_denoising}a and \ref{fig::results_denoising}b, where the three components are the closest to each other, is to be blamed for the underperformance. 
Since spectrogram zeros can only appear between components as the result of destructive interference \cite{flandrin2018explorations}, when three components (or more) are too close to each other, the lateral ones limit the repulsion of the zeros ``away'' from the component in the middle.
Consequently, the triangles over the components in between are not as elongated as those in other parts of the signal where the modes are further away from each other, making their detection more difficult.
Notice that such effect takes place for large SNRs, since low SNRs reduce the effective support of the signal in the plane (as seen in Fig. \ref{fig::shrinking_domains}), lessening the interference between signal components.

Fig. \ref{fig::error_dt}a shows an example of this situation, where the extraction mask of the DT method is depicted to illustrate why the method fails to recover the signal when the three components are close to each other.

Figures \ref{fig::error_dt}b and \ref{fig::error_dt}c show this in more detail, where it is possible to see how the triangles covering the ridges corresponding to the middle components are smaller than those corresponding to the uppermost and lowermost components.
Nevertheless, it is true that, in order to select those smaller triangles, one can still reduce $\ell_{\max}$. 
However, doing so results in less noise filtered out.
In conclusion, a complete disentanglement between signal and noise is not possible for these signals based solely on the length of the edges of the triangles. 
The same observations can be extrapolated to the Empty Space method, by considering the circumcircles of the Delaunay triangles rather than the triangles themselves.
This illustrates an interesting limitation of the most simple zero-based methods, and calls for a specific treatment of high-interference regions.

\begin{table}
\begin{tabular}{|c|c|c|c|} \hline
      \multirow{2}{*}{Method}  & \multicolumn{3}{c|}{Execution Time} \\ \cline{2-4}
                          & $N=512$ & $N=1024$  & $N=2048$    \\ \hline
       SST+RD             &  0.20   & 0.87   &  3.32    \\ \hline
       Contours           &  6.44  & 25.64  &  106.59    \\ \hline
       DT                 &  0.37   & 1.18  &  4.40    \\ \hline
       ES                 &  2.43   & 10.48  & 47.86     \\ \hline
       DT (Adaptive)      &  11.78  & 11.43  & 15.02   \\ \hline
       ES (Adaptive)      &  13.63  & 23.06  & 57.44   \\ \hline       
       PB                 &  0.60   & 3.62  & 21.88   \\ \hline
       SZC                &  5.20  & 20.01  &  138.00  \\ \hline
       Thresholding       &  0.18   & 0.44   &  1.29    \\ \hline
       
    \end{tabular}
  \caption{Average execution time of the methods compared in Benchmark 2 (in seconds). In the case of DT and ES, both the times using Algorithm \ref{alg::adapt_r0} (Adaptive) and with a fixed threshold. The execution time was computed for the signal used in Fig. \ref{fig::results_denoising}c, for $N\in\{512,1024,2048\}$. All the experiments were run on an Intel CORE i7 with 16 GB of RAM using Matlab 2022a and Python 3.10. }\label{tab:times}
\end{table}

\begin{figure}
\centering
    \setlength{\tabcolsep}{0.2pt}
    \renewcommand{\arraystretch}{0}
    \begin{tabular}{ccc}
    \multicolumn{3}{c}{\includegraphics[width=0.48\textwidth]{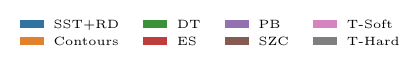}}\\
\includegraphics{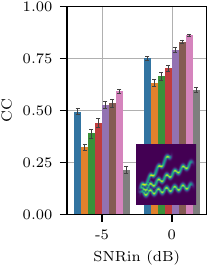} & \includegraphics{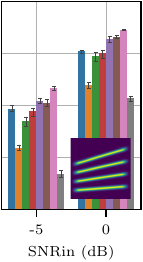}  
& \includegraphics{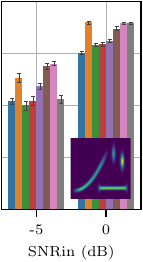} \\
(a)&(b)&(c) \\
    \end{tabular}
    \caption{Correlation coefficient for selected signals and SNRs. Results shown for the PB method were obtained using $\alpha=\beta=0.4$.} \label{fig:cc_results}
\end{figure}

In contrast, the SZC approach can deal with this cases since it considers the Delaunay triangles associated with SS zeros, regardless of the length of their edges.
Table \ref{tab:times} shows that this is achieved, however, at an elevated computational cost.
The SZC approach is not as scalable as the DT or ES approaches, with execution times at least one order of magnitude greater for longer signals ($N=2048$ for example).
This elevated computational cost is due to the simulations needed to compute the 2D histograms of zeros used in this method \cite{miramont2023unsupervised}.

Finally, Figs. \ref{fig::results_denoising}e and \ref{fig::results_denoising}f show that zero-based methods thrive for signals with impulsive transients or with a more extravagant circular component produced by a Hermite function \cite{flandrin2015time, meignen2016adaptive, flandrin2018explorations}, mainly because these approaches are inherently independent of the geometry of the signal domain.
This latter might as well constitute the main advantage of the methods based on spectrogram zeros.

\subsubsection{Ridge-based methods}
Ridge-detection based methods, like SST+RD and PB, can be expected to provide satisfactory results even when components are close to each other, as long as some degree of separation exists \cite{delprat1997global}.
Also, as shown by Fig. \ref{fig::results_denoising}, some of these methods, like SST+RD and Contours, keep a good performance when SNR decreases.
In particular, it can be seen that the Contours approach appears to provide a good performance for all the signals evaluated and for the lowest SNRs considered ($-5$ and $0$ dB).

However, using the QRF for low SNRs can be difficult to interpret and might lead to wrong conclusions.
In particular, if the recovered signal is identically equal to zero, the QRF would be also $0$ dB, which, for negative SNRs, might be considered as an improvement in the signal quality.

To complement what is shown in Fig. \ref{fig::results_denoising}, Figs. \ref{fig:cc_results}a-c display the correlation coefficient (CC):
\begin{equation}
 \operatorname{CC} =  \frac{\left \langle s, \tilde{s} \right\rangle}{\norm{s}_{2}\norm{\tilde{s}}_{2}},
\end{equation}
between the recovered signal ($\tilde{s}$) and the original signal ($s$), focusing on the SNRs of $-5$ and $0$ dB.
These complementary results were obtained using MCSM-Benchs to run the same benchmark but with a user-given performance metric.
It can be seen that, from the perspective of the CC, SST+RD yields better results, recovering a signal that is more correlated with the original one than the Contours approach.
 
Fig. \ref{fig::results_denoising}d shows that when the signal has components with fast amplitude modulation, the performance decreases, mainly because some parts of the ridges are not recovered.
The presence of impulse-like transients is also a limitation for some of these methods.
For example, Fig. \ref{fig::results_denoising}e describes a case where SST+RD and PB (for both parameter combinations) are less efficient than the Contours method, or the zero-based approaches, due to the impulsive like components that can be seen in the upper, rightmost part of the spectrogram of the signal.
Fig. \ref{fig:cc_results}c shows that Contours also obtains good results from the CC perspective for this signal.

Nevertheless, Table \ref{tab:times} shows that the execution time of the Contours approach is considerably larger than that of the other ridge-based approaches (SST+RD and PB).
This is caused by the thorough processing needed to estimate the projection angle and the signal support as the basins of attraction used in Contours approach (see Sec. \ref{sec:method_contours}).
In contrast, SST+RD and PB simply estimate the signal domain as a strip centered at the detected ridges.

\subsubsection{Hard and soft thresholding}
Similarly to the approaches based on the spectrogram zeros, the thresholding methods are also independent of the geometry of the signal domain, since they are built on assumptions on the noise rather than on the signal. 
As can be seen through all the panels in Fig. \ref{fig::results_denoising}, T-Soft and T-Hard perform better than the other approaches when the SNR is low, i.e. $0$ and $-5$ dB.
However, when the signal heavily deviates from the noise-only case, finding a satisfactory threshold becomes more difficult.
Since the thresholds are based on an estimation of the standard deviation of the (white Gaussian) noise given in Eq. \eqref{e:ht_estimator}, the more components the signal has, the more overestimated the threshold is, requiring a fine-tuning correction \cite{laurent2023novel}.

\begin{figure}[b]
  \centering
  \renewcommand{\arraystretch}{1}
  \begin{tabular}{cc}
  \multicolumn{2}{c}{\includegraphics[width = 0.48\textwidth]{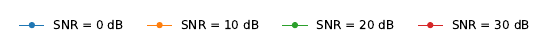}}\\
    \includegraphics{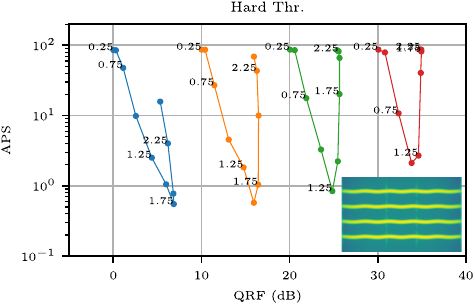} &
  \includegraphics{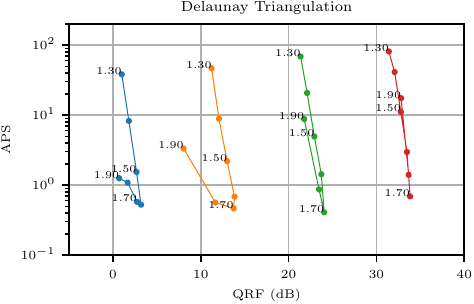} \\
    (a) & (b) \\
  \end{tabular}
  \caption{Contrasting two performance metrics (APS vs. QRF) for hard-thresholding and the Delaunay triangulation (indicated in the title of each plot). Each point represent the average over $50$ realizations. For the DT method, the value of $\ell_{\max}$ used is shown next to each point.  For Hard Thresholding, the value of $\lambda$ used to compute the threshold is shown next to each point.}\label{fig::benchmark3}
\end{figure}

\subsection{Benchmark 3. Signal reconstruction and perceptual quality evaluation.} \label{sec:bench3}

We now focus on the results of the third and last benchmark, that studied the trade-off between signal reconstruction and the introduction of perceptually unpleasant artifacts, hereafter termed musical noise, introduced by the Hard Thresholding and the DT methods.

Fig. \ref{fig::benchmark3} shows plots of the APS versus QRF for the two approaches and several SNRs. 
The spectrogram of the synthetic signal used in this case is shown in Fig. \ref{fig::benchmark3}a.

For both methods, low thresholds introduce almost no musical noise, since barely any coefficient is suppressed from the representations, hence the high value of APS and low QRF.

As the threshold is increased in both approaches, the number of isolated regions in the TF plane grows, leading first to an increase of the amount of musical noise, reflected as a decrease in APS. 
Meanwhile, QRF increases, since mostly noise-related coefficients are suppressed from the STFT.
This trade-off continues until APS reaches a minimum.

From this point on, the behavior of both methods differs. For higher thresholds, in the case of Hard Thresholding, the APS rapidly increases without incurring a great loss in QRF.
In contrast, for the DT method, especially for low SNRs ($0$ and $10$ dB), once the APS minimum is reached, it begins to grow again but at the expense of the QRF.
This implies that more musical noise is introduced by the DT method than by the, in principle equivalent, thresholding approach.

\begin{figure*}
    \centering
    \begin{tabular}{cc}
        \multicolumn{2}{c}{\includegraphics[width = 0.6\textwidth]{figures/legend_mn.pdf}} \\
        \includegraphics[width = 0.48\textwidth]{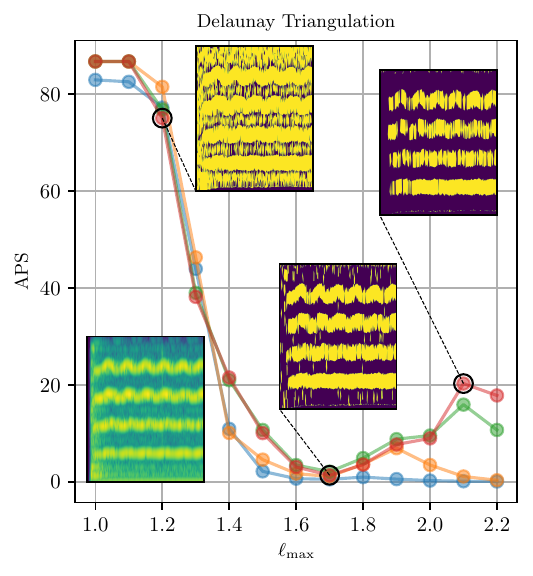} &
        \includegraphics[width = 0.48\textwidth]{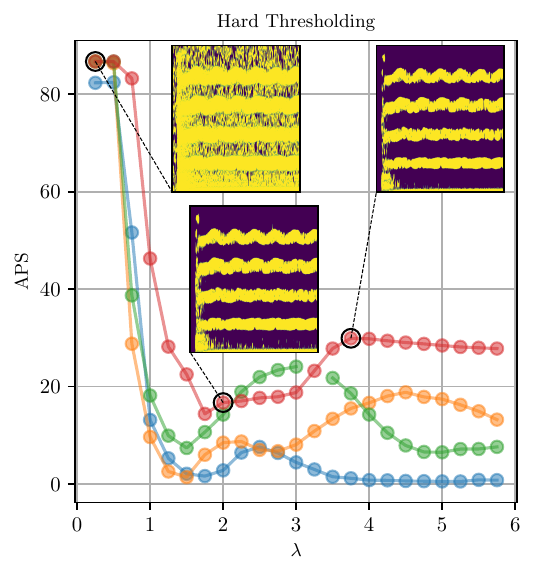} \\
    \end{tabular}

    \caption{Artifact-related perceptual score (APS) for a cello recording excerpt. (a) APS vs. $\ell_{\max}$ for the DT method. (b) APS vs. $\lambda$ for the Hard Thresholding approach. The spectrogram of the signal is shown in the bottom left corner of (a). Three extraction 1/0 masks obtain with the corresponding method are shown in (a) and (b) for $\text{SNR}=30$ dB and different values of the studied parameter.  }
    \label{fig:aps_cello}
\end{figure*}

To complement these results, Fig. \ref{fig:aps_cello} shows the values of APS obtained after applying the DT method (Fig. \ref{fig:aps_cello}a) and Hard Thresholding (Fig. \ref{fig:aps_cello}b) to a real-world audio signal, the spectrogram of which is shown in Fig. \ref{fig:aps_cello}a. 
The APS was computed after applying both methods for a range of parameters and SNRs.
It can be seen that, considering the same SNR, Hard Thresholding reaches a higher value of APS.
Fig. \ref{fig:aps_cello} also shows the extractions masks obtained for the minimum APS, and for the maximum APS obtained after the minimum.
The masks obtained with Hard Thresholding have a lower amount of isolated components, explaining the larger APS values obtained for this approach.

\section{Discussion} \label{sec:discussion}

\subsection{Signal detection tests based on the spectrogram zeros}
Considering the results from the first benchmark shown in Sec. \ref{sec:results:b1}, 
global tests (both MAD and rank tests) seem to be more suitable for signal detection, requiring fewer parameters than their point-wise counterparts for a similar performance.
The highest detection power was obtained by the global rank envelope method, particularly when the domain of $r$ is reduced to $[0.65;1.05]$.
For this test, there seems to be no statistically significant difference between using either $F$ or $\widetilde{F}$ for the detection task.
Based on this, using $\widetilde{F}$ would be preferable for signal detection, considering that one could later obtain an estimation of $r_{0}$ as a by-product, as shown in Algorithm \ref{alg::adapt_r0}.
Notice that using $\widetilde{F}$ increased the performance of most of the tests when compared to the use of $F$ only, showing that choosing an adequate summary function can booster the detection power of the tests.
Future work will evaluate other summary functions and add additional testing methods to increase the value of this benchmark.

Another point to consider is the number of simulations used in the MC tests.
In order to simplify the interpretation of the results, $2499$ simulations were used for all tests.
As mentioned before, this is the recommended number of simulations for global rank tests \cite{myllymaki2017global}. 
Yet, further experiments showed that the MAD or point-wise tests can perform well with less simulations.
Even as few as $199$ simulations are enough to obtain similar results to those shown in Fig. \ref{fig::power_test} for those tests.

\subsection{Signal estimation Based on the Spectrogram Zeros}
From the results shown before, it is clear that the most desirable quality of zero-based methods is their independence from the shape of $\mathcal{D}_{s}$.
This makes these approaches able to cope with Dirac impulses and with strongly modulated signals, producing adaptive estimations of the signal domain instead of fixed-width ribbons around the ridges.
A possible limitation, however, is set by the component separation in the TF plane.
When three or more components close to each other interfere, they can produce regions without zeros with a small area, similar to those one would expect in the noise-only case.
Both DT and ES methods studied here search for empty, larger-than-expected regions in the TF plane. Therefore, a situation like this makes these methods lose important parts of the signal.
This, however, is not a limitation for the more computationally expensive SZC approach.
Moreover, DT and ES  appear to be less effective than the methods based on the large values of the spectrogram for low SNR (for example $0$ and $-5$ dB).
This does not necessarily mean that zero-based methods cannot deliver in low SNR scenarios, since SZC provides good results in such case. Nevertheless, the latter uses an estimation of the variance of the noise in order to accurately classify the zeros, which makes it a kind of hybrid approach, i.e. not solely based on the location of zeros but also harnessing some extra information.

More work focusing on studying the robustness of the structure of the zeros of the spectrogram might provide new insights on how to improve these methods for low SNR cases, while keeping scalability of the computational cost.

The strategy introduced in Sec. \ref{sec:method_es} to estimate the parameters of ES and DT from the analyzed signal proved to be useful in the application of these techniques for different SNRs.
There are, however, some possible limitations.
First, because the detection test used in Algorithm \ref{alg::adapt_r0} is global, when the null hypothesis is rejected, it does not give information on the value of $r$ that led to that outcome.
Yet, the proposed estimation strategy appeared as an effective heuristic to determine $r_{0}$, particularly when $S = \widetilde{F}$.
However, as seen in Fig. \ref{fig::envelopes}b, the estimation obtained by means of Algorithm \ref{alg::adapt_r0}, although close to the optimal value of $r_{0}$, is still suboptimal.
The use of other functional statistics in future works might provide a better approximation of $r_{0}$.

Another limitation can be found for several components with different amplitude modulations.
The methods based on the spectrogram zeros could benefit from using a \emph{local} scale, i.e. dependent on the location in the TF plane, rather than one global scale of interaction given by $r_{0}$.

Moreover, considering the results shown in Sec. \ref{sec:bench3}, one must pay special attention when using the DT method for denoising in the case of audio signals, since it might introduce more undesired artifacts than traditional threshold-based methods.

Notice that some relations between the DT method and musical noise have been the subject of previous work \cite{hamon2017assessment}, aiming to quantify the amount of musical noise present in a signal. 
Although further exploration of this aspect is out the scope of this paper, a description of the isolated regions of the plane responsible of the musical noise, in terms of the Delaunay triangles, could be useful for post-processing techniques that would reduce the described artifacts. 
 
In summary, zero-based methods can have some general limitations.
When used to detect individual components, e.g. component retrieval or component counting, certain distance between them is necessary to avoid mode-mixture, i.e. considering two modes as one \cite{miramont2023eusipco, miramont2023unsupervised}.
Although this is also true for ridge-based methods, it seems that the minimum mode separation needs to be larger for zero-based approaches.
In terms of denoising, these methods appear to be more prone to introduce artifacts producing the perceptual degradation of the signal than other spectrogram thresholding techniques. 
 
\subsection{Signal estimation based on large values of the spectrogram}
As noticed before, methods based on the spectrogram zeros do not seem robust in low SNR cases, while methods based on spectrogram thresholding were the most efficient in such scenarios (observe, for example, T-Soft for $\text{SNRin}=-5$ dB in all the panels of Fig. \ref{fig::results_denoising}).
We were interested here in comparing the approaches described in Sec. \ref{sec:methods} in terms of their ability to estimate $\mathcal{D}_{s}$ and, consequently, disentangle signal from noise. 
Note that, unlike the rest of the explored methods, thresholding strategies are not able to identify individual components from the signal, which could be useful in other applications.

For higher SNRs, the Contours method explored here provided very satisfactory results, being able to extract impulsive-like components, and to estimate a variety of signal domain shapes.
These strengths are also shared by the zero-based methods, as mentioned before.
In fact, Contours is also an hybrid approach, as is SZC, since it uses the spectrogram zeros to segment the ridges, and similarly, this approach can be computationally expensive.
In spite of these similarities between SZC and Contours, the presence of interference between close components can be a hindrance for the former.
In this case, ridge detection-based approaches should be the methods of choice, more suited to extract individual components as long as the ridges are not superimposed or produce so-called \emph{time-frequency bubbles} \cite{delprat1997global}.

\subsection{General remarks}
Although only methods that work in the TF domain were compared in this article, MCSM-Benchs and the introduced  public benchmarks accept any signal denoising or detection approach, regardless of its nature. 
The only conditions methods need to fulfill is receiving a noisy signal as an input and delivering either a denoised version as an output (for the denoising methods), or a boolean indicating a signal is present (for detection approaches). 
For instance, time-domain based approaches \cite{harmouche2017sliding} will be included in the future in order to expand the collection of evaluated methods.

It should be noticed that, although only results for some signals provided by the \verb|SignalBank| class and real white Gaussian noise were reported, real-life signals and \emph{colored} noise sources can be incorporated if provided by the user.
Noise with non-white spectra requires a more nuanced analysis, since many methods would require an additional parameter tuning.
To simplify the analysis here, the parameter space of each approach was explored, in order to find values that resulted in the best possible performance for the range of SNRs and for the signals studied.
When the sensitivity to the parameters made this choice difficult, as with the PB method, two combination of parameters were used. 
In the case of the zero-based approaches, DT and ES methods automatically adapted to the signal using the strategy introduced in Sec. \ref{sec:method_dt}, whereas for SZC, the default parameters were used \cite{miramont2023unsupervised}.
Regarding the ridge-based approaches (Contours, PB and SST+RD), the number of components present in the signal was passed as an input parameter.
In a more realistic set-up, this would require preprocessing the signal in order to count the number of components \cite{sucic2011estimating}.

Finally, more results are shown in the online repository, as well as supplementary material for this paper.
Interested readers can interact with the Python notebooks made available in the repository to further explore the methods.

\section{Conclusion} \label{sec:conclusion}
In this paper, benchmarks of methods based on spectrogram zeros were used with the aim of exploring the performance for signal detection and signal estimation under stationary noise conditions.
In particular, situations where zero-based methods need further improvement were pointed out. 

The main conclusions of this work are:
\begin{itemize}
  \setlength\itemsep{0.1em}
    \item global signal detection tests based on restricted intervals for $r_0$ and variance-stabilized summary functions seem to be more powerful than previously proposed point-wise detection tests.
    \item the variance-stabilized empty-space function provides a more accurate estimation of $r_0$, which can be used to make zero-based methods adaptive.
    \item the more basic zero-based methods fail to deliver in low SNR conditions, but more complex approaches, like SZC, yielded better results at the expense of some computational burden.
    \item zero-based methods like DT seem to introduce more perceptually unpleasant artifacts than more traditional spectral subtraction techniques in audio applications.
\end{itemize}
All the methods described in this paper are now part of our public benchmarks, and we encourage interested researchers to upload their own algorithms to expand the library of available methods.

\section*{Acknowledgments}
This work was supported by grant ERC-2019-STG-851866, and French projects ANR20-CHIA-0002 and ASCETE-ANR19-CE48-0001.



\bibliographystyle{plain}


\begin{thebibliography}{10}

\bibitem{auger1995improving}
Fran{\c{c}}ois Auger and Patrick Flandrin.
\newblock Improving the readability of time-frequency and time-scale representations by the reassignment method.
\newblock {\em IEEE Transactions on {S}ignal {P}rocessing}, 43(5):1068--1089, 1995.

\bibitem{baddeley2014tests}
Adrian Baddeley, Peter~J Diggle, Andrew Hardegen, Thomas Lawrence, Robin~K Milne, and Gopalan Nair.
\newblock On tests of spatial pattern based on simulation envelopes.
\newblock {\em Ecological Monographs}, 84(3):477--489, 2014.

\bibitem{baddeley2015spatial}
Adrian Baddeley, Ege Rubak, and Rolf Turner.
\newblock {\em Spatial point patterns: methodology and applications with R}.
\newblock CRC press, 2015.

\bibitem{bardenet2018zeros}
R{\'e}mi Bardenet, Julien Flamant, and Pierre Chainais.
\newblock On the zeros of the spectrogram of white noise.
\newblock {\em Applied and Computational Harmonic Analysis}, 48(2):682--705, 2020.

\bibitem{bardenet2021time}
R{\'e}mi Bardenet and Adrien Hardy.
\newblock Time-frequency transforms of white noises and {G}aussian analytic functions.
\newblock {\em Applied and computational harmonic analysis}, 50:73--104, 2021.

\bibitem{bartz2020benchmarking}
Thomas Bartz-Beielstein, Carola Doerr, Daan van~den Berg, Jakob Bossek, Sowmya Chandrasekaran, Tome Eftimov, Andreas Fischbach, Pascal Kerschke, William La~Cava, Manuel Lopez-Ibanez, et~al.
\newblock Benchmarking in optimization: Best practice and open issues.
\newblock {\em arXiv preprint arXiv:2007.03488}, 2020.

\bibitem{carmona1999multiridge}
Ren{\'e}~A Carmona, Wen~L Hwang, and Bruno Torr{\'e}sani.
\newblock Multiridge detection and time-frequency reconstruction.
\newblock {\em IEEE transactions on signal processing}, 47(2):480--492, 1999.

\bibitem{chassande1997differential}
E~Chassande-Mottin, Ingrid Daubechies, Francois Auger, and Patrick Flandrin.
\newblock Differential reassignment.
\newblock {\em IEEE {S}ignal {P}rocessing {L}etters}, 4(10):293--294, 1997.

\bibitem{cohen1966generalized}
Leon Cohen.
\newblock Generalized phase-space distribution functions.
\newblock {\em Journal of Mathematical Physics}, 7(5):781--786, 1966.

\bibitem{colominas2020fully}
Marcelo~A Colominas, Sylvain Meignen, and Duong-Hung Pham.
\newblock Fully adaptive ridge detection based on {STFT} phase information.
\newblock {\em IEEE Signal Processing Letters}, 27:620--624, 2020.

\bibitem{courbot2023sparse}
Jean-Baptiste Courbot, Ali Moukadem, Bruno Colicchio, and Alain Dieterlen.
\newblock Sparse off-the-grid computation of the zeros of stft.
\newblock {\em IEEE Signal Processing Letters}, 2023.

\bibitem{daubechies2011synchrosqueezed}
Ingrid Daubechies, Jianfeng Lu, and Hau-Tieng Wu.
\newblock Synchrosqueezed wavelet transforms: An empirical mode decomposition-like tool.
\newblock {\em Applied and computational harmonic analysis}, 30(2):243--261, 2011.

\bibitem{daubechies2017nonlinear}
Ingrid Daubechies and Stephane Maes.
\newblock A nonlinear squeezing of the continuous wavelet transform based on auditory nerve models.
\newblock In {\em Wavelets in medicine and biology}, pages 527--546. Routledge, 2017.

\bibitem{delprat1997global}
Nathalie Delprat.
\newblock Global frequency modulation laws extraction from the gabor transform of a signal: A first study of the interacting components case.
\newblock {\em IEEE Transactions on Speech and Audio Processing}, 5(1):64--71, 1997.

\bibitem{donoho1995noising}
David~L Donoho.
\newblock De-noising by soft-thresholding.
\newblock {\em IEEE {T}ransactions on {I}nformation {T}heory}, 41(3):613--627, 1995.

\bibitem{donoho1994ideal}
David~L Donoho and Jain~M Johnstone.
\newblock Ideal spatial adaptation by wavelet shrinkage.
\newblock {\em Biometrika}, 81(3):425--455, 1994.

\bibitem{emiya2011subjective}
Valentin Emiya, Emmanuel Vincent, Niklas Harlander, and Volker Hohmann.
\newblock Subjective and objective quality assessment of audio source separation.
\newblock {\em IEEE Transactions on Audio, Speech, and Language Processing}, 19(7):2046--2057, 2011.

\bibitem{flandrin1998time}
Patrick Flandrin.
\newblock {\em Time-frequency/time-scale analysis}.
\newblock Academic press, 1998.

\bibitem{flandrin2015time}
Patrick Flandrin.
\newblock Time--frequency filtering based on spectrogram zeros.
\newblock {\em IEEE Signal Processing Letters}, 22(11):2137--2141, 2015.

\bibitem{flandrin2018explorations}
Patrick Flandrin.
\newblock {\em Explorations in time-frequency analysis}.
\newblock Cambridge University Press, 2018.

\bibitem{gao1998wavelet}
Hong-Ye Gao.
\newblock Wavelet shrinkage denoising using the non-negative {G}arrote.
\newblock {\em Journal of Computational and Graphical Statistics}, 7(4):469--488, 1998.

\bibitem{ghosh2022signal}
Subhroshekhar Ghosh, Meixia Lin, and Dongfang Sun.
\newblock Signal analysis via the stochastic geometry of spectrogram level sets.
\newblock {\em IEEE Transactions on Signal Processing}, 70:1104--1117, 2022.

\bibitem{goh1998postprocessing}
Zenton Goh, Kah-Chye Tan, and TG~Tan.
\newblock Postprocessing method for suppressing musical noise generated by spectral subtraction.
\newblock {\em IEEE Transactions on Speech and Audio Processing}, 6(3):287--292, 1998.

\bibitem{hamon2017assessment}
Ronan Hamon, Valentin Emiya, Lucas Rencker, Wenwu Wang, and Mark Plumbley.
\newblock Assessment of musical noise using localization of isolated peaks in time-frequency domain.
\newblock In {\em 2017 IEEE International Conference on Acoustics, Speech and Signal Processing (ICASSP)}, pages 696--700. IEEE, 2017.

\bibitem{hansen2021coco}
N.~Hansen, A.~Auger, R.~Ros, O.~Mersmann, T.~Tu{\v s}ar, and D.~Brockhoff.
\newblock {COCO}: A platform for comparing continuous optimizers in a black-box setting.
\newblock {\em Optimization Methods and Software}, 36:114--144, 2021.

\bibitem{harmouche2017sliding}
Jinane Harmouche, Dominique Fourer, Fran{\c{c}}ois Auger, Pierre Borgnat, and Patrick Flandrin.
\newblock The sliding singular spectrum analysis: A data-driven nonstationary signal decomposition tool.
\newblock {\em IEEE Transactions on Signal Processing}, 66(1):251--263, 2017.

\bibitem{HOUZ10}
H.~Holden, B.~{\O}ksendal, J.~Ub{\o}e, and T.~Zhang.
\newblock {\em Stochastic partial differential equations}.
\newblock Springer, second edition, 2010.

\bibitem{hough2009zeros}
John~Ben Hough, Manjunath Krishnapur, Yuval Peres, et~al.
\newblock {\em Zeros of Gaussian analytic functions and determinantal point processes}, volume~51.
\newblock American Mathematical Soc., 2009.

\bibitem{miramont2023eusipco}
D.~Fourer J.~M.~Miramont, Q.~Legros and F.~Auger.
\newblock Benchmarks of multi-component signal analysis methods.
\newblock In {\em EUSIPCO}, 2023.

\bibitem{kay2006fundamentals}
SM~Kay.
\newblock Fundamentals of statistical signal processing volume i, and ii: Estimation theory and detection theory.
\newblock {\em Beijin: Publishing House of Electronics Industry}, 41, 2006.

\bibitem{koliander2019filtering}
G{\"u}nther Koliander, Luis~Daniel Abreu, Antti Haimi, and Jos{\'e}~Luis Romero.
\newblock Filtering the continuous wavelet transform using hyperbolic triangulations.
\newblock In {\em 2019 13th International conference on Sampling Theory and Applications (SampTA)}, pages 1--4. IEEE, 2019.

\bibitem{laurent2023novel}
N~Laurent, S~Meignen, MA~Colominas, JM~Miramont, and F~Auger.
\newblock A novel approach based on vorono{\"\i} cells to classify spectrogram zeros of multicomponent signals.
\newblock In {\em ICASSP 2023-2023 IEEE International Conference on Acoustics, Speech and Signal Processing (ICASSP)}, pages 1--5. IEEE, 2023.

\bibitem{laurent2020novel}
Nils Laurent and Sylvain Meignen.
\newblock A novel time-frequency technique for mode retrieval based on linear chirp approximation.
\newblock {\em IEEE Signal Processing Letters}, 27:935--939, 2020.

\bibitem{laurent2021novel}
Nils Laurent and Sylvain Meignen.
\newblock A novel ridge detector for nonstationary multicomponent signals: Development and application to robust mode retrieval.
\newblock {\em IEEE Transactions on Signal Processing}, 69:3325--3336, 2021.

\bibitem{legros2021novel}
Quentin Legros and Dominique Fourer.
\newblock A novel pseudo-bayesian approach for robust multi-ridge detection and mode retrieval.
\newblock In {\em 2021 29th European Signal Processing Conference (EUSIPCO)}, pages 1925--1929. IEEE, 2021.

\bibitem{legros2022pb}
Quentin Legros and Dominique Fourer.
\newblock Pseudo-bayesian approach for robust mode detection and extraction based on the stft.
\newblock {\em Sensors}, 23(1):85, Dec 2022.

\bibitem{legros2022time}
Quentin Legros and Dominique Fourer.
\newblock Time-frequency ridge estimation of multi-component signals using sparse modeling of signal innovation.
\newblock {\em arXiv preprint arXiv:2212.11343}, 2022.

\bibitem{legros2022instantaneous}
Quentin Legros, Dominique Fourer, Sylvain Meignen, and Marcelo~A Colominas.
\newblock Instantaneous frequency estimation in multi-component signals using stochastic {EM} algorithm.
\newblock {\em arXiv preprint arXiv:2203.16334}, 2022.

\bibitem{lim2012sparse}
Yoonseob Lim, Barbara Shinn-Cunningham, and Timothy~J Gardner.
\newblock Sparse contour representations of sound.
\newblock {\em IEEE Signal Processing Letters}, 19(10):684--687, 2012.

\bibitem{mallat2008tour}
St{\'e}phane Mallat.
\newblock {\em A wavelet tour of signal processing, Third Edition: The Sparse Way}.
\newblock Academic Press, 2008.

\bibitem{mattson2020mlperf}
Peter Mattson, Christine Cheng, Gregory Diamos, Cody Coleman, Paulius Micikevicius, David Patterson, Hanlin Tang, Gu-Yeon Wei, Peter Bailis, Victor Bittorf, et~al.
\newblock Mlperf training benchmark.
\newblock {\em Proceedings of Machine Learning and Systems}, 2:336--349, 2020.

\bibitem{meignen2016adaptive}
Sylvain Meignen, Thomas Oberlin, Philippe Depalle, Patrick Flandrin, and Stephen McLaughlin.
\newblock Adaptive multimode signal reconstruction from time--frequency representations.
\newblock {\em Philosophical Transactions of the Royal Society A: Mathematical, Physical and Engineering Sciences}, 374(2065):20150205, 2016.

\bibitem{meignen2018retrieval}
Sylvain Meignen and Duong-Hung Pham.
\newblock Retrieval of the modes of multicomponent signals from downsampled short-time fourier transform.
\newblock {\em IEEE Transactions on Signal Processing}, 66(23):6204--6215, 2018.

\bibitem{michelsanti2021overview}
Daniel Michelsanti, Zheng-Hua Tan, Shi-Xiong Zhang, Yong Xu, Meng Yu, Dong Yu, and Jesper Jensen.
\newblock An overview of deep-learning-based audio-visual speech enhancement and separation.
\newblock {\em IEEE/ACM Transactions on Audio, Speech, and Language Processing}, 29:1368--1396, 2021.

\bibitem{miramont2023unsupervised}
Juan~M Miramont, Fran{\c{c}}ois Auger, Marcelo~A Colominas, Nils Laurent, and Sylvain Meignen.
\newblock Unsupervised classification of the spectrogram zeros with an application to signal detection and denoising.
\newblock {\em Signal Processing}, page 109250, 2023.

\bibitem{miramont2022public}
Juan~M Miramont, R{\'e}mi Bardenet, Pierre Chainais, and Fran{\c{c}}ois Auger.
\newblock A public benchmark for denoising and detection methods.
\newblock In {\em XXVIII{\`e}me Colloque Francophone du GRETSI}, pages 1--4, 2022.

\bibitem{benchopt}
Thomas Moreau, Mathurin Massias, Alexandre Gramfort, Pierre Ablin, Pierre-Antoine Bannier, Benjamin Charlier, Mathieu Dagréou, Tom Dupré~la Tour, Ghislain Durif, Cassio F.~Dantas, Quentin Klopfenstein, Johan Larsson, En~Lai, Tanguy Lefort, Benoit Malézieux, Badr Moufad, Binh T.~Nguyen, Alain Rakotomamonjy, Zaccharie Ramzi, Joseph Salmon, and Samuel Vaiter.
\newblock Benchopt: Reproducible, efficient and collaborative optimization benchmarks.
\newblock In {\em NeurIPS}, 2022.

\bibitem{moukadem2021zeros}
Ali Moukadem, Jean-Baptiste Courbot, Bruno Colicchio, and Alain Dieterlen.
\newblock {On the Zeros of the Stockwell and Morlet Wavelet Transforms}.
\newblock working paper or preprint, June 2021.

\bibitem{myllymaki2017global}
Mari Myllym{\"a}ki, Tom{\'a}{\v{s}} Mrkvi{\v{c}}ka, Pavel Grabarnik, Henri Seijo, and Ute Hahn.
\newblock Global envelope tests for spatial processes.
\newblock {\em Journal of the Royal Statistical Society: Series B (Statistical Methodology)}, 79(2):381--404, 2017.

\bibitem{najari2022neyman}
Nader Najari, Mohammad~Q Vahidi~Asl, and Abdollah Jalilian.
\newblock {N}eyman-{S}cott process with skew-normal clusters.
\newblock {\em Communications in Statistics-Theory and Methods}, 51(14):4692--4711, 2022.

\bibitem{villasana2023eusipco}
L.~Villemoes P.~J.~Villasana, J.~Klejsa and F.~Auger.
\newblock Distribution preserving source separation with time frequency predictive models.
\newblock In {\em EUSIPCO}, 2023.

\bibitem{pascal2022covariant}
Barbara Pascal and R{\'e}mi Bardenet.
\newblock A covariant, discrete time-frequency representation tailored for zero-based signal detection.
\newblock {\em IEEE Transactions on Signal Processing}, 2022.

\bibitem{pascalfamille}
Barbara Pascar and R{\'e}mi Bardenet.
\newblock Une famille de repr{\'e}sentations covariantes de signaux discrets et son applicationa la d{\'e}tection de signauxa partir de leurs z{\'e}ros.
\newblock In {\em XXVIII{\`e}me Colloque Francophone du GRETSI}, pages 1--4, 2022.

\bibitem{pham2017adaptive}
Duong-Hung Pham and Sylvain Meignen.
\newblock An adaptive computation of contour representations for mode decomposition.
\newblock {\em IEEE Signal Processing Letters}, 24(11):1596--1600, 2017.

\bibitem{sawada2019}
Hiroshi Sawada, Nobutaka Ono, Hirokazu Kameoka, Daichi Kitamura, and Hiroshi Saruwatari.
\newblock A review of blind source separation methods: two converging routes to ilrma originating from ica and nmf.
\newblock {\em APSIPA Transactions on Signal and Information Processing}, 8:e12, 2019.

\bibitem{simonetta2019multimodal}
Federico Simonetta, Stavros Ntalampiras, and Federico Avanzini.
\newblock Multimodal music information processing and retrieval: Survey and future challenges.
\newblock In {\em 2019 international workshop on multilayer music representation and processing (MMRP)}, pages 10--18. IEEE, 2019.

\bibitem{sucic2011estimating}
Victor Sucic, Nicoletta Saulig, and Boualem Boashash.
\newblock Estimating the number of components of a multicomponent nonstationary signal using the short-term time-frequency r{\'e}nyi entropy.
\newblock {\em EURASIP Journal on Advances in Signal Processing}, 2011:1--11, 2011.

\bibitem{thakur2013synchrosqueezing}
Gaurav Thakur, Eugene Brevdo, Neven~S Fu{\v{c}}kar, and Hau-Tieng Wu.
\newblock The synchrosqueezing algorithm for time-varying spectral analysis: Robustness properties and new paleoclimate applications.
\newblock {\em Signal Processing}, 93(5):1079--1094, 2013.

\bibitem{pandas2}
{The Pandas Development Team}.
\newblock pandas-dev/pandas.
\newblock {\em Zenodo}, Sep 2022.

\bibitem{torcoli2019improved}
Matteo Torcoli.
\newblock An improved measure of musical noise based on spectral kurtosis.
\newblock In {\em 2019 IEEE Workshop on Applications of Signal Processing to Audio and Acoustics (WASPAA)}, pages 90--94. IEEE, 2019.

\bibitem{torcoli2018comparing}
Matteo Torcoli and Sascha Dick.
\newblock Comparing the effect of audio coding artifacts on objective quality measures and on subjective ratings.
\newblock In {\em Audio Engineering Society Convention 144}. Audio Engineering Society, 2018.

\bibitem{torcoli2021objective}
Matteo Torcoli, Thorsten Kastner, and J{\"u}rgen Herre.
\newblock Objective measures of perceptual audio quality reviewed: An evaluation of their application domain dependence.
\newblock {\em IEEE/ACM Transactions on Audio, Speech, and Language Processing}, 29:1530--1541, 2021.

\bibitem{van2004detection}
Harry~L Van~Trees.
\newblock {\em Detection, estimation, and modulation theory, part I: detection, estimation, and linear modulation theory}.
\newblock John Wiley \& Sons, 2004.

\bibitem{vincent2012improved}
Emmanuel Vincent.
\newblock Improved perceptual metrics for the evaluation of audio source separation.
\newblock In {\em Latent Variable Analysis and Signal Separation: 10th International Conference, LVA/ICA 2012, Tel Aviv, Israel, March 12-15, 2012. Proceedings 10}, pages 430--437. Springer, 2012.

\bibitem{pandas1}
{W}es {M}c{K}inney.
\newblock {D}ata {S}tructures for {S}tatistical {C}omputing in {P}ython.
\newblock In {S}t\'efan van~der {W}alt and {J}arrod {M}illman, editors, {\em {P}roceedings of the 9th {P}ython in {S}cience {C}onference}, pages 56 -- 61, 2010.

\bibitem{whalen2013detection}
Anthony~D Whalen.
\newblock {\em Detection of signals in noise}.
\newblock Academic press, 2013.

\end{thebibliography}





\end{document}